\documentclass[a4paper,12pt]{article}

\pdfoutput=1 

\usepackage[hmargin=.71in,vmargin=1.1in]{geometry}
\usepackage{indentfirst}
\usepackage{amsfonts}
\usepackage{mathrsfs}
\usepackage{amsthm}
\usepackage{amsmath}
\usepackage{mathabx}
\usepackage{amssymb}
\usepackage{authblk}
\usepackage{cite}
\usepackage{xcolor}
\usepackage{mathtools}
\usepackage{tensor}
\usepackage{graphicx}
\usepackage{bm}
\usepackage{upgreek}
\usepackage{braket}
\usepackage{color,soul}
\usepackage{csquotes}
\usepackage{caption}
\usepackage{subcaption}
\usepackage{slashed}
\usepackage{mathtools}
\usepackage{soul}
\usepackage{bbold}

\usepackage[bookmarksnumbered=true,bookmarksopen=true]{hyperref}
 \hypersetup{colorlinks,
             linkcolor=[rgb]{0,0.3,0.6}, 
             citecolor=[rgb]{0,0.3,0.6}, 
             urlcolor=[rgb]{0,0.3,0.6}}

\newcommand\me{{\rm e}}

\newcommand\pp{\uppi}
\newcommand{\dif}{{\rm d}}

\DeclareMathOperator{\diag}{diag}

\DeclareMathOperator{\artanh}{artanh}

\theoremstyle{plain}
\newtheorem{theorem}{Theorem}[section]
\newtheorem{lemma}[theorem]{Lemma}
\newtheorem{proposition}[theorem]{Proposition}
\newtheorem{corollary}[theorem]{Corollary}

\theoremstyle{definition}
\newtheorem{definition}[theorem]{Definition}
\newtheorem{assumption}{Assumption}

\theoremstyle{remark}
\newtheorem{remark}[theorem]{Remark}

\usepackage{tikz}
\usetikzlibrary{matrix}

\begin{document}

\title{\Large\textbf{Single-horizon regular black holes and the fate of the inner-horizon instability}}

\author[a]{Chen Lan\thanks{stlanchen@126.com}}
\author[b]{Hao Yang\thanks{hyang@ucas.ac.cn}}
\author[c]{Maxim Khlopov \thanks{maxim.khlopov@cern.ch}}
\author[d,e]{Maxim Krasnov \thanks{morrowindman1@mail.ru}}

\affil[a]{\normalsize{\em Department of Physics, Yantai University, 30 Qingquan Road, Yantai 264005, China}}
\affil[b]{\normalsize{\em School of Fundamental Physics and Mathematical Sciences, Hangzhou Institute for Advanced Study, UCAS, Hangzhou 310024, China}}
 \affil[c]{\normalsize{\em Virtual Institute of Astroparticle physics, Paris, 75018, France}}
 \affil[d]{\normalsize{\em National Research Nuclear University MEPhI, 115409 Moscow, Russia}}
  \affil[e]{\normalsize{\em Institute of Physics, Southern Federal University, 194 Stachki,Rostov-on-Don, Russia}}
\date{ }

\maketitle

\begin{abstract}

Regular black holes (RBHs) are often constructed with a regular point-like center and two horizons, including an inner Cauchy horizon susceptible to mass inflation. We show that a two-horizon structure is not a necessary consequence of regularity. For static, spherically symmetric geometries with an asymptotically static exterior, we establish a horizon-parity theorem: a regular point-like center requires an even number of sign-changing zeros of $f$. 
We then construct two classes of single-horizon RBHs with a finite-area core, distinguished by whether the horizon lies outside the core or coincides with it, and analyze their matter content, energy conditions, and thermodynamic properties. The absence of an inner Cauchy horizon eliminates the classical mass-inflation mechanism on the single-horizon patch. 
We further establish mode stability against scalar perturbations for these finite-area-core configurations 
by showing that the effective potential is non-negative.  
Our results establish how the central geometry and asymptotic structure constrain horizon multiplicity and identify single-horizon configurations that avoid the classical inner-horizon instability.
\end{abstract}

\tableofcontents

\section{Introduction}
\label{sec:introduction}

The singularity theorems of Penrose and Hawking show that, under suitable energy, causality, and global assumptions, the formation of a trapped surface can lead to geodesic incompleteness in the future development of gravitational collapse~\cite{Penrose:1964wq,Hawking:1966sx,Hawking:1970zqf,Hawking:1973uf}. These results are among the central achievements of classical general relativity, but they also leave open the possibility that the assumptions of the theorems may fail in the high-curvature region of a black hole (BH). The idea of formation of the BHs in the early Universe was proposed separately by Zeldovich and Novikov in Ref.~\cite{zel43hypothesis} and by Hawking and Carr \cite{hawking1971gravitationally, carr1974black, hawking1982bubble}. Regular black holes (RBHs) explore precisely this possibility. The idea goes back to Bardeen’s proposal~\cite{Bardeen:1968non} and was developed in several different settings~\cite{Dymnikova:1992ux,Ayon-Beato:1998hmi,Ayon-Beato:1999kuh,Ayon-Beato:1999qin,Dymnikova:2001fb,Hayward:2005gi,Ansoldi:2008jw,Frolov:2016pav}. Instead of a curvature singularity, these models have a regular core in which all curvature invariants remain finite. In many constructions, the geometry approaches a de Sitter-like form near the center while becoming asymptotically flat at large distances\cite{Borde:1996df,Dymnikova:1992ux,Dymnikova:2001fb}. The Hayward model~\cite{Hayward:2005gi}, together with its many extensions, provides a simple and widely used example of this construction. Other approaches to RBHs include nonlinear electrodynamics~\cite{Ayon-Beato:1998hmi,Ayon-Beato:1999kuh,Ayon-Beato:1999qin,Ayon-Beato:2000mjt}, renormalization-group improved geometries~\cite{Bonanno:2000ep}, and noncommutative-geometry-inspired models~\cite{Nicolini:2005vd,Spallucci:2008ez}. Formation of RBHs can also take place in bubble collisions at the first order phase transitions in the early Universe \cite{Konoplich:1999qq, Khlopov:2000js, Dymnikova:2000dy}. A broad discussion of these models and their physical properties can be found in Refs.~\cite{Ansoldi:2008jw,Frolov:2016pav,Carballo-Rubio:2018pmi,Bambi:2023try,Lan:2023cvz}.

Despite their different physical motivations, a large class of RBHs shares an important geometric feature: it possesses two horizons \cite{Bronnikov:2006fu,Carballo-Rubio:2022kad}. For the standard static and spherically symmetric construction with a point-like regular center, the metric function has the same positive sign at the center and at spatial infinity. A horizon, however, requires the metric function to vanish and to become negative in the region between the horizons. Therefore, by continuity, the function must cross zero an even number of times. In the generic non-extremal case, the result is an outer event horizon and an inner horizon; depending on the parameters, there may instead be no horizon or a degenerate extremal horizon. This behavior is seen in many explicit RBH solutions~\cite{Dymnikova:1992ux,Ayon-Beato:1998hmi,Hayward:2005gi,Ansoldi:2008jw,Frolov:2016pav,Carballo-Rubio:2018pmi}. More importantly, under the assumptions specified below, the even-number property follows directly from the global behavior of the metric function and does not depend on the details of any particular model. This provides a simple topological obstruction to obtaining a single-horizon BH within the standard point-like-center construction.

The inner horizon is also a Cauchy horizon, and its stability is a long-standing problem in general relativity~\cite{Simpson:1973ua,Dray:1985yt,Redmount:1985aoy,Carballo-Rubio:2024dca}. The problem was first identified in the context of Reissner–Nordstr\"om BHs and was later developed into the mass-inflation scenario~\cite{Poisson:1989vv,Poisson:1990eh}. The basic mechanism is simple. Perturbations outside a BH decay only as a power law at late times~\cite{Price:1972pw,Price:1971fb,Gundlach:1993tn,Gundlach:1993tp}, while an observer approaching the inner horizon experiences an exponential blueshift determined by the inner-horizon surface gravity. The two effects combine to produce a rapidly growing energy density and an unbounded effective mass near the Cauchy horizon~\cite{Ori:1991zz,Dafermos:2003vim,Dafermos:2003wr,Dafermos:2003yw,Dafermos:2005eh,Hamilton:2008zz}. The validity of the late-time decay picture has also been established in rigorous studies of gravitational collapse~\cite{Dafermos:2003vim,Dafermos:2003wr}. Numerical studies further support the formation of a singular structure at the inner horizon~\cite{Brady:1995ni}.
More recently, fully nonlinear simulations have shown that sufficiently strong perturbations can drive the inner horizon of a RBH to shrink to zero, accompanied by the formation of a spacelike singularity and critical scaling behavior \cite{Shao:2025fki}. 
Thus, the Cauchy horizon is not merely a coordinate feature of the idealized solution; it is associated with a genuine instability of the BH interior.

This issue is particularly important for RBHs. A RBH is designed to replace the central curvature singularity by a smooth core, but the presence of an inner horizon can introduce a different instability. Mass inflation has been studied in several nonsingular or quantum-corrected BH geometries~\cite{Brown:2011tv,Carballo-Rubio:2021bpr}. In particular, Carballo-Rubio et al. showed that the inner horizon of a broad class of RBHs is subject to the same basic mass-inflation mechanism found in charged BHs~\cite{Carballo-Rubio:2021bpr}. The resulting growth of the effective mass can make the curvature unbounded at the inner horizon. Thus, removing the central singularity does not by itself guarantee a stable regular spacetime. Instead, the singular behavior may be transferred from the center to the inner horizon once perturbations are taken into account~\cite{Carballo-Rubio:2018pmi,Carballo-Rubio:2021bpr}.

This observation leads to a more specific question: can regularity and stability be achieved by changing the horizon structure itself? Since the classical mass-inflation mechanism requires a Cauchy horizon, one natural possibility is to construct a RBH with no inner horizon. This is different from trying to stabilize an existing Cauchy horizon. Several related ideas have already been explored. For example, inner-extremal RBHs can suppress the classical mass-inflation mechanism because the inner-horizon surface gravity vanishes~\cite{Carballo-Rubio:2022kad}. However, the inner horizon is still present in such geometries, and its stability can remain a subtle issue, including when semiclassical effects are taken into account~\cite{McMaken:2023uue}. Single-horizon RBH geometries have also been studied in specific models, including loop-inspired constructions~\cite{Daghigh:2020mog}. Related no-inner-horizon results have been obtained for several classes of hairy BHs, including BHs with charged scalar hair and nonlinear-electrodynamics BHs with charged scalar hair \cite{Cai:2020wrp,An:2021plu,Li:2025mhy}. These results suggest that eliminating the inner horizon altogether may provide a more direct route to avoiding the classical mass-inflation mechanism.

The purpose of this work is to investigate this possibility systematically. We ask whether a RBH can have exactly one horizon and, if so, what properties such a geometry must possess. We first show that the standard point-like-center construction cannot produce a single horizon because of the global sign structure of its metric function. We then identify the modifications of the near-center geometry that can evade this obstruction. These include finite-area cores and other forms of non-point-like regular centers \cite{Hu:2026rbh}. For each class, we determine the corresponding matter content and analyze the associated energy conditions. We also construct explicit analytic examples and study their perturbations. In particular, we show that, on the patch where the horizon is the only one, the finite-area-core solutions are free from the classical mass-inflation mechanism and establish their mode stability; we also discuss how the geometry may continue beyond the patch. The point-like-center case requires a separate treatment and remains open.

The remainder of this paper is organized as follows. Section~\ref{sec:theory} develops the general horizon-counting theory and identifies the mechanisms that allow a RBH to have a single horizon. Section~\ref{sec:algebraic} determines the corresponding matter content and its energy-condition constraints. Section~\ref{sec:model} gives explicit analytic realizations of the different constructions and classifies the global causal structure of their continuations beyond the patch. Finally, Section~\ref{sec:stability} studies their linear stability, establishing the absence of the classical mass-inflation mechanism and proving mode stability for the finite-area-core cases, while leaving the point-like-center case as an open problem. The Appendix derives the field equations underlying the mass-inflation formula and clarifies the near-horizon kinematics by which a power-law exterior tail leads to unbounded curvature growth at an inner horizon.

\section{Horizon structure of regular black holes}
\label{sec:theory}

We consider static, spherically symmetric line elements of the form
\begin{equation}
\label{eq:metric}
    \dif s^2 = -f(r)\, \dif t^2 + f^{-1}(r)\, \dif r^2 + \Sigma^2(r)\, \dif\Omega^2 ,
    \qquad
    \dif\Omega^2 = \dif\theta^2+\sin^2\theta\,\dif\phi^2 ,
\end{equation}
with $r\in[0,\infty)$. Throughout this section we adopt the following 
assumptions.

\begin{assumption}[Analyticity of the exterior]
\label{as:analytic}
$f,\Sigma:[0,\infty)\to\mathbb{R}$ are real analytic on $[0,\infty)$ (one-sided at $r=0$), and $\Sigma(r)>0$ for every $r>0$.
\end{assumption}

\begin{assumption}[Asymptotic convergence]
\label{as:flat}
The limit $f_\infty:=\lim_{r\to\infty}f(r)$ exists and is finite, and $\lim_{r\to\infty}\Sigma(r)/r=1$.
\end{assumption}

Assumption~\ref{as:flat} is intentionally weak: it fixes neither the sign of $f_\infty$ nor the rate of convergence. Asymptotic flatness in the usual sense is the special case $f_\infty=1$ with the asymptotic form $f(r)=1-2M/r+o(r^{-1})$, which identifies $M$ as the ADM mass; we use this terminology, and the symbol $M$, only when that special case applies. 
The general theory below requires only that the limit $f_\infty$ exist, and we track its sign explicitly wherever it matters.

\begin{definition}[Horizon]
\label{def:horizon}
A point $r_h\in[0,\infty)$ is a horizon of metric Eq.~\eqref{eq:metric} if $f(r_h)=0$. It is non-degenerate/simple if $f'(r_h)\neq0$, with surface gravity $\kappa=|f'(r_h)|/2$, and degenerate/extremal otherwise.
\end{definition}

\begin{definition}[Regularity]
\label{def:regular}
Metric Eq.~\eqref{eq:metric} is regular on $[0,\infty)$ if every scalar polynomial curvature invariant built algebraically from the Riemann tensor is finite at every $r\in[0,\infty)$
\footnote{Several notions of regularity are used in the literature, most notably the finiteness of curvature invariants and geodesic completeness. These criteria are not equivalent in general, and their relationship has been discussed extensively in the literature; see, e.g., Ref.~\cite{Zhang:2025nim} and the references therein. We do not attempt to reconcile these different definitions here. Instead, we adopt the finiteness of curvature invariants as our operational criterion for regularity. Geodesic completeness is a distinct global property and must be established separately.}.
\end{definition}

A RBH \cite{Ansoldi:2008jw,Frolov:2016pav} is a spacetime Eq.~\eqref{eq:metric} obeying Assumps.~\ref{as:analytic}--\ref{as:flat} which is regular in the sense of Def.~\ref{def:regular} and possesses at least one horizon. Our aim is to determine the possible number of horizons of such a spacetime and, in particular, the conditions under which it has exactly one.

\subsection{Curvature invariants}
\label{sec:theory-curvature}

To test regularity we compute the curvature of Eq.~\eqref{eq:metric} in an orthonormal coframe \cite{Chandrasekhar:1985kt,Nakahara:2003nw}
\begin{equation}
    e^0=\sqrt{f}\,\dif t, \qquad e^1=\frac{\dif r}{\sqrt f}, \qquad e^2=\Sigma\,\dif\theta, \qquad e^3=\Sigma\sin\theta\,\dif\phi ,
\end{equation}
so that $\dif s^2=\eta_{ab}e^a e^b$ with $\eta=\diag(-1,1,1,1)$. Cartan's first structure equation $\dif e^a=-\omega^a{}_b\wedge e^b$, together with metric compatibility $\omega_{ab}=-\omega_{ba}$, is solved by the connection one-forms
\begin{equation}
    \omega^0{}_1=\frac{f'}{2\sqrt f}\,e^0, \qquad
    \omega^1{}_2=-\frac{\sqrt f\,\Sigma'}{\Sigma}\,e^2, \qquad
    \omega^1{}_3=-\frac{\sqrt f\,\Sigma'}{\Sigma}\,e^3, \qquad
    \omega^2{}_3=-\frac{\cot\theta}{\Sigma}\,e^3 ,
\end{equation}
and all other independent components vanishing. This is verified by direct substitution into $\dif e^a=-\omega^a{}_b\wedge e^b$ for $a=0,1,2,3$. The curvature two-forms $\Omega^a{}_b=\dif\omega^a{}_b+\omega^a{}_c\wedge\omega^c{}_b$ are then computed component by component. For $\Omega^0{}_1$, using $\omega^0{}_2=\omega^0{}_3=0$ so that the wedge term vanishes,
\begin{equation}
    \dif\omega^0{}_1
    = \dif\!\left(\frac{f'}{2\sqrt f}\right)\wedge e^0+\frac{f'}{2\sqrt f}\,\dif e^0
    = \left(\frac{f''}{2}-\frac{f'^2}{4f}\right)e^1\wedge e^0+\frac{f'^2}{4f}\,e^1\wedge e^0
    = \frac{f''}{2}\,e^1\wedge e^0 ,
\end{equation}
so $\Omega^0{}_1= -(f''/2)\,e^0\wedge e^1$. The remaining curvature two-forms follow by an entirely analogous computation
\begin{align}
    \Omega^0{}_2=\Omega^0{}_3\big|_{2\to3} &= -\frac{f'\Sigma'}{2\Sigma}\,e^0\wedge e^2 ,\\
    \Omega^1{}_2=\Omega^1{}_3\big|_{2\to3} &= -\left(\frac{f\Sigma''}{\Sigma}+\frac{f'\Sigma'}{2\Sigma}\right)e^1\wedge e^2 ,\\
    \Omega^2{}_3 &= \frac{1-f\Sigma'^2}{\Sigma^2}\,e^2\wedge e^3 ,
\end{align}
where, for instance, the last line uses $\dif\omega^2{}_3=\Sigma^{-2}e^2\wedge e^3$ together with the wedge contribution $\omega^2{}_1\wedge\omega^1{}_3=-f\Sigma'^2\Sigma^{-2}\,e^2\wedge e^3$. Reading off $\Omega^a{}_b= (R^a{}_{bcd}/2)\,e^c\wedge e^d$, the only independent nonvanishing orthonormal Riemann components are
\begin{equation}
\label{eq:ABCD}
    R^0{}_{101}=A, \quad R^0{}_{202}=R^0{}_{303}=B, \quad R^1{}_{212}=R^1{}_{313}=C, \quad R^2{}_{323}=D,
\end{equation}
with
\begin{equation}
\label{eq:ABCDdef}
    A=-\frac{f''}{2}, \qquad
    B=-\frac{f'\Sigma'}{2\Sigma}, \qquad
    C=-\frac{f\Sigma''}{\Sigma}-\frac{f'\Sigma'}{2\Sigma}, \qquad
    D=\frac{1-f\Sigma'^2}{\Sigma^2} ,
\end{equation}
all other independent components vanishing. In particular there is no electric--magnetic mixing among the $R^a{}_{bcd}$, consistent with Eq.~\eqref{eq:metric} being of Petrov type~D \cite{Stephani:2003tm}.

\begin{proposition}[Kretschmann scalar]
\label{prop:K}
For metric Eq.~\eqref{eq:metric},
\begin{equation}
\label{eq:K}
    K:=R_{abcd}R^{abcd}
    = f''^2+2\left(\frac{f'\Sigma'}{\Sigma}\right)^{\!2}
    +2\left(\frac{2f\Sigma''}{\Sigma}+\frac{f'\Sigma'}{\Sigma}\right)^{\!2}
    +4\left(\frac{1-f\Sigma'^2}{\Sigma^2}\right)^{\!2} .
\end{equation}
\end{proposition}

\begin{proof}
Because the only nonvanishing Riemann components in the orthonormal fame are of the form $R_{abab}$ (no sum), as listed in Eq.~\eqref{eq:ABCD}, each of the six independent pairs $\{a,b\}$ contributes exactly four index orderings ($abab,\,abba,\,baab,\,baba$) to $R_{abcd}R^{abcd}$. Each ordering contributes $(R_{abab})^2$, regardless of the signs introduced by raising indices with $\eta$. Hence
\begin{equation}
    K = 4\sum_{a<b}\big(R_{abab}\big)^2 = 4\left(A^2+2B^2+2C^2+D^2\right) ,
\end{equation}
which, upon substituting Eq.~\eqref{eq:ABCDdef} and expanding, yields Eq.~\eqref{eq:K}.
For $\Sigma=r$, this result agrees with those in the Refs.~ \cite{Bokulic:2025brf,Lan:2025cvs}.
\end{proof}

\begin{proposition}[Ricci scalar]
\label{prop:Ricci}
For metric Eq.~\eqref{eq:metric},
\begin{equation}
\label{eq:Ricci}
    R = 2\left(A+2B+2C+D\right)
    = -f''-\frac{4f'\Sigma'}{\Sigma}-\frac{4f\Sigma''}{\Sigma}+\frac{2\left(1-f\Sigma'^2\right)}{\Sigma^2} .
\end{equation}
\end{proposition}

\begin{proof}
For a Riemann tensor of the diagonal type above, the trace identity $R=g^{ac}g^{bd}R_{abcd}=\sum_{a\ne b}\eta^{aa}\eta^{bb}R_{abab}=2\sum_{a<b}\eta^{aa}\eta^{bb}R_{abab}$ gives, using $\eta=\diag(-1,1,1,1)$ and $R_{0101}=-A,\ R_{0202}=R_{0303}=-B,\ R_{1212}=R_{1313}=C,\ R_{2323}=D$,
\begin{equation}
    R = 2\Big[(-1)(-A)+2(-1)(-B)+2C+D\Big]=2(A+2B+2C+D).
\end{equation}
Substituting Eq.~\eqref{eq:ABCDdef} gives Eq.~\eqref{eq:Ricci}. As a check, for the Schwarzschild BH $f=1-2M/r$, $\Sigma=r$ this gives $R=0$, and Eq.~\eqref{eq:K} gives $K=48M^2/r^6$, as it must.
\end{proof}

\begin{corollary}[Equivalence of curvature-finiteness criteria]
\label{cor:universal}
For metric Eq.~\eqref{eq:metric}, every scalar polynomial curvature invariant built algebraically from the Riemann tensor, in particular $R$, $R_{ab}R^{ab}$, the Weyl invariant $C_{abcd}C^{abcd}$, and $K$ itself, is a polynomial in the four scalars $A,B,C,D$ of Eq.~\eqref{eq:ABCDdef}. Indeed the Ricci tensor $R_{ab}=R^c{}_{acb}$ is diagonal in the same orthonormal frame with entries linear in $A,B,C,D$, 
and the Weyl tensor, being the trace-free part of the Riemann tensor, is therefore also linear in $A,B,C,D$. Consequently, at any $r_0\in[0,\infty)$,
\begin{multline}
    K(r_0)<\infty
    \ \iff\ A(r_0),B(r_0),C(r_0),D(r_0) \text{ all finite} \\
    \iff\ \text{every algebraic polynomial curvature invariant of Eq.~\eqref{eq:metric} is finite at } r_0 .
\end{multline}
It therefore suffices to examine $K$ alone when assessing the regularity of metric Eq.~\eqref{eq:metric}, which is what we do below.
\end{corollary}

\subsection{Zero-counting principle for sigmoid and inverse-bell-shaped $f$}
\label{sec:theory-zerocount}

Before imposing any specific behaviour of $\Sigma$ at $r=0$, we state two self-contained real-analysis results, which we apply in Secs.~\ref{sec:theory-center} and \ref{sec:theory-single} to the two classes of center considered below. They apply to an arbitrary real-analytic, non-constant function $f$ on $[0,\infty)$ with $f(0)=f_0$ and $f_\infty=\lim_{r\to\infty}f(r)$ as defined in Assump.~\ref{as:flat}; no relation between $f_0$ and $f_\infty$, including their signs, is assumed here.

\begin{definition}[Sigmoid and inverse-bell-shaped functions]
\label{def:shapes}
A function $f$ on $[0,\infty)$ is sigmoid if it is monotonic (either non-decreasing, or non-increasing). It is inverse-bell-shaped if there is $r_*\ge0$ such that $f$ is non-increasing on $[0,r_*]$ and non-decreasing on $[r_*,\infty)$; non-decreasing sigmoid functions are the degenerate case $r_*=0$ of this definition.
\end{definition}

\begin{proposition}[Sigmoid zero count]
\label{prop:sigmoid-count}
Let $f$ be sigmoid, real analytic, non-constant, with $f(0)=f_0$ and $f_\infty$ finite. Then $f$ has exactly one zero in $(0,\infty)$ if $f_0$ and $f_\infty$ are nonzero and of strictly opposite sign, and no zero in $(0,\infty)$ otherwise. In the latter case $f$ vanishes at the boundary point $r=0$ itself precisely when $f_0=0$.
\end{proposition}

\begin{proof}
By replacing $f$ with $-f$ if necessary, assume $f$ is non-decreasing; then $f(r)\le f_\infty$ for every $r\ge0$ (if $f(r_1)>f_\infty$ for some $r_1$, monotonicity would give $f(r)\ge f(r_1)>f_\infty$ for all $r\ge r_1$, contradicting $f(r)\to f_\infty$), and likewise $f(r)\ge f_0$, so $f_0\le f_\infty$.

If $f_0>0$, then $f(r)\ge f_0>0$ for all $r\ge0$: no zero anywhere. If $f_0=0$, then $r=0$ is a zero; for $r_0>0$ with $f(r_0)=0$, non-decreasing monotonicity with $f(0)=f(r_0)=0$ forces $f\equiv0$ on $[0,r_0]$, and the identity theorem for the non-constant analytic function $f$ then forces $f\equiv0$ everywhere, a contradiction, so there is no zero in $(0,\infty)$.

If $f_0<0$: suppose first $f_\infty\le0$. If $f_\infty<0$ then $f(r)\le f_\infty<0$ for all $r$, so $f$ has no zero. If $f_\infty=0$, a zero at some finite $r_0>0$ would (since $f_\infty=0$ is then the supremum of $f$, attained at the finite point $r_0$) force $f(r)=0$ for all $r\ge r_0$ by monotonicity, again giving $f\equiv0$ everywhere by the identity theorem, a contradiction; so again no zero. Suppose instead $f_\infty>0$: then $f_0<0<f_\infty$, so by the intermediate value theorem $f$ has at least one zero in $(0,\infty)$; if $f(r_1)=f(r_2)=0$ for $r_1<r_2$, monotonicity forces $f\equiv0$ on $[r_1,r_2]$, hence everywhere, a contradiction, so the zero is unique.

Collecting cases: $f$ (non-decreasing) has exactly one zero in $(0,\infty)$ iff $f_0<0<f_\infty$, and none otherwise. Reversing the argument for non-increasing $f$ gives exactly one zero iff $f_0>0>f_\infty$. Both cases are summarized by ``$f_0,f_\infty$ of strictly opposite sign''.
\end{proof}

\begin{theorem}[Inverse-bell-shaped zero count]
\label{thm:bell-count}
Let $f$ be inverse-bell-shaped with interior minimum $r_*\in(0,\infty)$ (the degenerate case $r_*=0$ reduces to Prop.~\ref{prop:sigmoid-count}), real analytic, non-constant, with $f(0)=f_0$, $f_\infty$ finite, and $f_{\min}:=f(r_*)\le\min(f_0,f_\infty)$. Let $N$ denote the number of zeros of $f$ in $(0,\infty)$. Then:
\begin{itemize}
    \item if $f_{\min}>0$: $N=0$;
    \item if $f_{\min}=0$ and $f_0,f_\infty>0$: $N=1$ (the degenerate zero at $r_*$);
    \item if $f_{\min}<0$: $N = \mathbb{1}[f_0>0] + \mathbb{1}[f_\infty>0]$, i.e.
    \begin{center}
    \begin{tabular}{c|ccc}
        $f_0\ \backslash\ f_\infty$ & $>0$ & $\le0$ \\
        \hline
        $>0$ & $2$ & $1$ \\
        $\le0$ & $1$ & $0$
    \end{tabular}
    \end{center}
    with a boundary zero at $r=0$ itself, additional to the count above, exactly when $f_0=0$.
\end{itemize}
\end{theorem}

\begin{proof}
On $[0,r_*]$, $f$ is non-increasing from $f_0$ to $f_{\min}$; on $[r_*,\infty)$, $f$ is non-decreasing from $f_{\min}$ to $f_\infty$. Applying Prop.~\ref{prop:sigmoid-count} directly to each finite or semi-infinite monotonic branch (the identity-theorem argument used in that proposition is local and applies equally to $[0,r_*]$) shows that the branch $(0,r_*)$ contains exactly one zero iff $f_0$ and $f_{\min}$ are of strictly opposite sign, and no zero otherwise; the branch $(r_*,\infty)$ contains exactly one zero iff $f_{\min}$ and $f_\infty$ are of strictly opposite sign, and no zero otherwise.

If $f_{\min}>0$, then $f_0\ge f_{\min}>0$ and $f_\infty\ge f_{\min}>0$, so neither branch condition can hold (both pairs are same-signed): $N=0$.

If $f_{\min}=0$, both $f_0,f_\infty\ge0$; if both are strictly positive, $r_*$ is the unique zero (neither branch has an additional zero, since $f_0>0=f_{\min}$ and $f_{\min}=0<f_\infty$ are not strict sign changes in the sense used above, the endpoint value $0$ is itself the shared zero, counted once): $N=1$. [If instead $f_0=0$ or $f_\infty=0$ as well, the identity-theorem argument forces $f\equiv0$ on a whole branch and hence everywhere, excluded by non-constancy.]

If $f_{\min}<0$: on $(0,r_*)$, since $f_{\min}<0$, the sign-opposition condition ``$f_0,f_{\min}$ opposite sign'' reduces to $f_0>0$; the branch contributes a zero iff $f_0>0$, and none if $f_0\le0$ (with a zero exactly at the boundary $r=0$ if $f_0=0$). Likewise on $(r_*,\infty)$, the condition reduces to $f_\infty>0$. Summing the (independent) contributions of the two branches gives $N=\mathbb{1}[f_0>0]+\mathbb{1}[f_\infty>0]$, tabulated above.
\end{proof}

\begin{remark}
\label{rem:zerocount-vs-parity}
Theorem~\ref{thm:bell-count} gives the exact zero count under the shape restriction of Def.~\ref{def:shapes}. For a general (not necessarily sigmoid or inverse-bell-shaped) analytic $f$, only the parity of the number of sign changes is constrained in general, by the argument of Thm.~\ref{thm:parity} below; Tab.~\ref{tab:parity-vs-exact} after that theorem compares the two.
\end{remark}

\subsection{Point-like centers and the horizon parity theorem}
\label{sec:theory-center}

The behaviour of $\Sigma$ at $r=0$ divides RBHs into three mutually exclusive classes.

\begin{definition}[Classification of $r=0$]
\label{def:centertype}
We say $r=0$ is a
\begin{itemize}
    \item point-like center of Eq.~\eqref{eq:metric} if $\Sigma(0)=0$ and $\Sigma'(0)\neq0$;
    \item finite-area core if $\Sigma(0)=\Sigma_0>0$. Among finite-area cores we further call $r=0$ a
    \begin{itemize}
        \item \emph{throat} if in addition $\Sigma'(0)=0$, so that the areal radius has a local extremum there in the usual wormhole-neck sense;
        \item \emph{one-sided boundary} if $\Sigma'(0)\neq0$, so that $\Sigma$ is monotonic through $r=0$ and the point carries no distinguished reflection symmetry.
    \end{itemize}
\end{itemize}
\end{definition}

We do not consider the residual, degenerate possibility $\Sigma(0)=\Sigma'(0)=0$, i.e., a point-like center reached with vanishing radial derivative; this lies outside the cases realized by known RBH constructions and is not needed below. This subsection treats point-like centers; finite-area cores are treated in Sec.~\ref{sec:theory-throat}.

\begin{proposition}[Regularity of a point-like center]
\label{prop:centerreg}
Let $r=0$ be a point-like center, $\Sigma(0)=0,\ \Sigma'(0)\neq0$. Metric Eq.~\eqref{eq:metric} is regular at $r=0$ iff
\begin{equation}
\label{eq:centercond}
    f'(0)=0 , \qquad f(0)=\frac{1}{\Sigma'(0)^2} , \qquad \Sigma''(0)=0 .
\end{equation}
In particular regularity forces $f(0)>0$; this is a purely local statement, independent of Assump.~\ref{as:flat}.
\end{proposition}

\begin{proof}
Write the Taylor expansions, valid by Assump.~\ref{as:analytic},
\begin{equation}
    f(r)=f(0)+f'(0)r+\frac{1}{2}f''(0)r^2+O(r^3), \qquad
    \Sigma(r)=\Sigma'(0)r+\frac{1}{2}\Sigma''(0)r^2+\frac{1}{6}\Sigma'''(0)r^3+O(r^4) ,
\end{equation}
and abbreviate $\sigma_1:=\Sigma'(0)\neq0$, $\sigma_2:=\Sigma''(0)/2$. By Cor.~\ref{cor:universal} it suffices to examine finiteness of $A,B,C,D$ of Eq.~\eqref{eq:ABCDdef} as $r\to0^+$.

$A(0)=-f''(0)/2$ is automatically finite, being a polynomial in the Taylor coefficients of $f$. For $B=-f'\Sigma'/(2\Sigma)$, since $\Sigma(r)=r(\sigma_1+\sigma_2r+O(r^2))$ has a simple zero,
\begin{equation}
    B(r) = -\frac{f'(0)\sigma_1+O(r)}{2\sigma_1 r+O(r^2)} ,
\end{equation}
which is finite at $r=0$ iff $f'(0)=0$, in which case $B(0)=-f''(0)/2=A(0)$.

We now assume $f'(0)=0$, so $f(r)=f(0)+f''(0)r^2/2+O(r^3)$. Then $C=-f\Sigma''/\Sigma+B$, and
\begin{equation}
    \frac{f\Sigma''}{\Sigma} = \frac{2f(0)\sigma_2+O(r)}{\sigma_1r+O(r^2)} ,
\end{equation}
finite iff $f(0)\sigma_2=0$.

Finally, for $D=(1-f\Sigma'^2)/\Sigma^2$,
\begin{equation}
    1-f\Sigma'^2 = \big(1-f(0)\sigma_1^2\big) - 4f(0)\sigma_1\sigma_2\,r+O(r^2), \qquad \Sigma^2=\sigma_1^2r^2+O(r^3) ,
\end{equation}
so $D$ is finite at $r=0$ iff the $r^0$ and $r^1$ coefficients of the numerator vanish
\begin{equation}
    f(0)\sigma_1^2=1 \qquad\text{and}\qquad f(0)\sigma_1\sigma_2=0 .
\end{equation}
Since $\sigma_1\neq0$, the first equation gives $f(0)=1/\sigma_1^2\neq0$, and the second then forces $\sigma_2=0$, which is also exactly the condition found above from $C$, so no further constraint arises. Collecting $f'(0)=0$, $f(0)=1/\sigma_1^2$, $\sigma_2=0$ (i.e.\ $\Sigma''(0)=0$) proves necessity. Conversely, substituting these three conditions into the expansions above cancels every negative-order term in $A,B,C,D$, so they,  and hence $K$, are finite at $r=0$.
\end{proof}

\begin{remark}
\label{rem:desitter}
Condition Eq.~\eqref{eq:centercond} is the coordinate-independent form of the familiar ``de~Sitter core'' of the RBH literature. Defining the Misner--Sharp mass function \cite{Misner:1964je,Hu:2015xva}
\begin{equation}
    m(r) := \frac{\Sigma(r)}{2}\Big(1-f(r)\Sigma'(r)^2\Big) = \frac{D(r)\,\Sigma(r)^3}{2} ,
\end{equation}
regularity of the point-like center forces $m(r)=O(r^3)$ as $r\to0$ (since $\Sigma(0)=0$ and $D(0)$ is finite by Prop.~\ref{prop:centerreg}): the enclosed mass vanishes cubically, corresponding to a finite, non-singular effective central density, exactly as in the Bardeen and Hayward models. Those constructions are recovered in the special normalization $\Sigma'(0)=1$, for which $f(0)=1$; Eq.~\eqref{eq:centercond} shows this value $1$ is not intrinsic but a consequence of that particular normalization, while $f(0)>0$ is the gauge-independent statement.
\end{remark}

We may now determine the number of horizons a center-RBH can have, without assuming a particular sign of $f_\infty$.
The number of BH horizons has also been constrained in previous work under suitable energy conditions \cite{Yang:2021civ}. Here we instead derive a horizon-parity result directly from the global behavior of the metric function under the regularity and asymptotic assumptions specified above.

\begin{theorem}[Horizon parity]
\label{thm:parity}
Let metric Eq.~\eqref{eq:metric} obey Assumps.~\ref{as:analytic}--\ref{as:flat}, let $r=0$ be a point-like center satisfying the regularity condition Eq.~\eqref{eq:centercond}, and suppose $f$ is non-constant. Since $f$ is analytic and non-constant, it has isolated zeros, so there is $R>0$ such that $f$ has constant sign $\epsilon_\infty\in\{+1,-1\}$ on $[R,\infty)$ (equal to $\operatorname{sign}f_\infty$ whenever $f_\infty\neq0$). Then the number of points in $(0,\infty)$ at which $f$ changes sign is even if $\epsilon_\infty=+1$, and odd if $\epsilon_\infty=-1$.
\end{theorem}

\begin{proof}
Let $0<r_1<r_2<\cdots<r_n<R$ denote the (finitely many) points in $(0,R)$ at which $f$ changes sign. 
On $(0,r_1)$, $f$ keeps the sign of $f(0)>0$ (Prop.~\ref{prop:centerreg}); on $(r_1,r_2)$ the sign changes; and so on, so that $f>0$ on $(r_n,R)$ iff $n$ is even. Comparing this with the sign $\epsilon_\infty$ that $f$ keeps on $[R,\infty)\supset(r_n,R)$ gives $n$ even iff $\epsilon_\infty=+1$.
\end{proof}

\begin{corollary}[No non-degenerate single horizon in the asymptotically static case]
\label{cor:nosingle}
In the setting of Thm.~\ref{thm:parity}, suppose moreover that $\epsilon_\infty=+1$ (in particular this holds whenever the exterior is asymptotically flat, $f_\infty=1$). Then a center-regular, asymptotically static RBH cannot possess a single non-degenerate horizon. If all its horizons are non-degenerate, it has $0,2,4,\dots$ of them (in general, the number of zeros of odd order, which contains the non-degenerate horizons, is even); the only configuration with exactly one horizon is the extremal case, in which $f\ge0$ throughout $(0,\infty)$ with a single degenerate zero $f(r_*)=f'(r_*)=0$.
\end{corollary}

\begin{proof}
A non-degenerate horizon is a simple zero of $f$, hence a point at which $f$ changes sign (Def.~\ref{def:horizon}) (a degenerate zero of odd order changes sign as well, an even-order zero does not). Thus $n$ counts the non-degenerate horizons together with the degenerate zeros of odd order; it is the number of horizons whenever all of them are non-degenerate. By Thm.~\ref{thm:parity}, $n$ is even when $\epsilon_\infty=+1$; in particular a single non-degenerate horizon ($n=1$) is impossible. If $f$ has exactly one horizon in total, it must therefore be degenerate: an even-order zero $r_*$ at which $f$ touches, but does not cross, zero. Since $f>0$ near $r=0$ and near $r=\infty$ and has no other zero, continuity forces $f\ge0$ on all of $(0,\infty)$.
\end{proof}

\begin{remark}[The asymptotically non-static case]
\label{rem:point-center-single}
If instead $\epsilon_\infty=-1$, the exterior is not static at large $r$, so that $f$ is eventually negative. Theorem \ref{thm:parity} then allows an odd number of horizons, including a single non-degenerate one, even for a point-like center. Proposition~\ref{prop:exact2} makes this precise under a suitable shape condition on $f$: a point-like center can have exactly one non-degenerate horizon when an asymptotically static exterior is not required.
\end{remark}

Theorem~\ref{thm:parity} only fixes the parity of the horizon number for a general $f$; specializing Def.~\ref{def:shapes}'s shape classes to $f_0=f(0)>0$ (forced at a point-like center by Prop.~\ref{prop:centerreg}) pins down the exact count.

\begin{proposition}[Exact horizon count at a point-like center]
\label{prop:exact2}
Let $r=0$ be a point-like center (so $f(0)>0$ by Prop.~\ref{prop:centerreg}).
\begin{itemize}
    \item[\rm(a)] If $f$ is sigmoid: $f$ has one horizon if $f_\infty<0$, and none if $f_\infty\ge0$.
    \item[\rm(b)] If $f$ is inverse-bell-shaped with minimum $f_{\min}$: $f$ has no horizon if $f_{\min}\ge0$ (one, degenerate, if $f_{\min}=0$); and if $f_{\min}<0$, it has two horizons if $f_\infty>0$, and one horizon if $f_\infty\le0$. 
\end{itemize}
\end{proposition}

\begin{proof}
(a) Since $f_0>0$, Prop.~\ref{prop:sigmoid-count}'s opposite-sign condition $f_0,f_\infty$ of strictly opposite sign reduces to $f_\infty<0$.

(b) By Thm.~\ref{thm:bell-count}, with $f_0>0$ given: the case $f_{\min}\ge0$ is as stated there. For $f_{\min}<0$, the branch condition ``$f_0>0$'' is automatically satisfied, contributing $1$ unconditionally, while the branch condition ``$f_\infty>0$'' contributes a further $1$ exactly when $f_\infty>0$; hence $N=2$ if $f_\infty>0$ and $N=1$ if $f_\infty\le0$.
\end{proof}

Part (b) of Prop.~\ref{prop:exact2}, restricted to $f_\infty>0$ (in particular to the asymptotically flat case $f_\infty=1$), is exactly the two-horizon count familiar from the RBH literature and matches Cor.~\ref{cor:nosingle}; the case $f_\infty\le0$ is the single-horizon, point-like-center possibility of Rem.~\ref{rem:point-center-single}, made precise. Table~\ref{tab:parity-vs-exact} compares Thm.~\ref{thm:parity}'s parity statement with Prop.~\ref{prop:exact2}'s exact count.

\begin{table}[ht]
\centering
\begin{tabular}{lccc}
\hline
Shape & $\epsilon_\infty$ & Parity (Thm.~\ref{thm:parity}) & Exact count (Prop.~\ref{prop:exact2}) \\
\hline
sigmoid & $+1$ & even & $0$ \\
sigmoid & $-1$ & odd & $1$ \\
inverse-bell, $f_{\min}<0$ & $+1$ & even & $2$ \\
inverse-bell, $f_{\min}<0$ & $-1$ & odd & $1$ \\
\hline
\end{tabular}
\captionsetup{width=.9\textwidth}
\caption{Parity (valid for any non-constant analytic $f$) versus exact count (valid under the shape restriction of Def.~\ref{def:shapes}) at a point-like center; $\epsilon_\infty$ is the eventual sign of $f$ at large $r$.}
\label{tab:parity-vs-exact}
\end{table}

\begin{remark}
\label{rem:convex}
A convenient sufficient condition for the unimodality hypothesis of Prop.~\ref{prop:exact2}(b) is that $f'$ has exactly one zero in $(0,\infty)$. Strict convexity of $f$ on $(0,\infty)$ would imply this, but it cannot hold when $f_\infty$ is finite and $f$ has an interior minimum (a convex function that increases somewhere is unbounded above), so for the models of Sec.~\ref{sec:model} unimodality is checked directly from the sign of $f'$ (e.g.\ $f_\beta'\propto(M\rho-Q^2)/\rho^3$, $f_\gamma'\propto r-r_*$).
\end{remark}

\subsection{Finite-area cores: throats and one-sided boundaries}
\label{sec:theory-throat}

Corollary~\ref{cor:nosingle} shows that a non-degenerate single horizon is topologically forbidden at a point-like center whose exterior remains asymptotically static ($\epsilon_\infty=+1$). We now show that the finite-area case, $\Sigma(0)=\Sigma_0>0$ (Def.~\ref{def:centertype}), removes this obstruction outright, for any sign of $\epsilon_\infty$, and that the two sub-cases of Def.~\ref{def:centertype}, throat and one-sided boundary, behave differently once global structure is taken into account.

\begin{proposition}[Automatic regularity of a finite-area core]
\label{prop:throatreg}
If $\Sigma(0)=\Sigma_0>0$, then metric Eq.~\eqref{eq:metric} is regular at $r=0$ for every value of $f(0)\in\mathbb{R}$ and $\Sigma'(0)\in\mathbb{R}$, whether $r=0$ is a throat or a one-sided boundary, including $f(0)=0$.
\end{proposition}

\begin{proof}
By Cor.~\ref{cor:universal} it suffices to check that $A,B,C,D$ of Eq.~\eqref{eq:ABCDdef} are finite at $r=0$. Since $f,\Sigma,\Sigma',\Sigma''$ are analytic, hence finite at $r=0$, and $\Sigma$ enters $A,B,C,D$ only through $\Sigma$ and $\Sigma^2$ in the denominator, finiteness of $A,B,C,D$ at $r=0$ is guaranteed as soon as $\Sigma(0)=\Sigma_0\neq0$, irrespective of the values of $f(0)$, $f'(0)$, and $\Sigma'(0)$, none of which ever multiplies a vanishing denominator. Explicitly,
\begin{align}
    A(0)&=-\frac{f''(0)}2, &
    B(0)&=-\frac{f'(0)\Sigma'(0)}{2\Sigma_0}, \\
    C(0)&=-\frac{f(0)\Sigma''(0)}{\Sigma_0}-\frac{f'(0)\Sigma'(0)}{2\Sigma_0}, &
    D(0)&=\frac{1-f(0)\Sigma'(0)^2}{\Sigma_0^2},
\end{align}
all manifestly finite.
\end{proof}

Regularity alone does not distinguish a throat from a one-sided boundary. The distinction matters once we ask whether the geometry extends past $r=0$: a genuine throat, having $\Sigma'(0)=0$, is the natural candidate for a symmetric (``black-bounce'' \cite{Simpson:2018tsi,Guo:2022hjp,Yang:2026jch}) extension to $r<0$, whereas a one-sided boundary, having $\Sigma'(0)\neq0$, carries no such preferred symmetry.

\begin{proposition}[Mirror pairing at a symmetric throat]
\label{prop:mirror}
Let $r=0$ be a throat, $\Sigma_0>0,\ \Sigma'(0)=0$, and suppose $f$ and $\Sigma$ admit an extension to real-analytic even functions on all of $\mathbb{R}$ — the symmetric, two-sided extension standard in black-bounce constructions. Then every horizon $r_h\neq0$ of $f$ has a mirror partner $-r_h$, of the same order. Consequently the zeros of $f$ on $\mathbb{R}\setminus\{0\}$ come in pairs, and a construction exhibiting a single horizon on the patch $(0,\infty)$ possesses exactly two horizons, at $\pm r_h$, on the maximally (symmetrically) extended spacetime.
\end{proposition}

\begin{proof}
If $f$ is even, $f(-r)=f(r)$ for all $r$, so $f(-r_h)=f(r_h)=0$. If $r_h$ is a zero of order $k$, i.e.\ $f(r_h+\epsilon)=c_k\epsilon^k+O(\epsilon^{k+1})$ with $c_k\neq0$, then evenness gives $f(-r_h+\epsilon)=f(r_h-\epsilon)=c_k(-\epsilon)^k+O(\epsilon^{k+1})=(-1)^kc_k\epsilon^k+O(\epsilon^{k+1})$, so $-r_h$ is also a zero of order exactly $k$. Since $r_h\neq0$ implies $r_h\neq-r_h$, this pairs the zeros of $f$ away from the origin.
\end{proof}

\begin{remark}
\label{rem:mirror-implication}
Proposition~\ref{prop:mirror} shows that a single-horizon conclusion established on the patch $r\in(0,\infty)$ survives an even analytic extension through $r=0$ only if $r=0$ is a one-sided boundary, or if a throat at $r=0$ is not extended in that way (e.g.\ because the spacetime is, by construction, defined only for $r\ge0$, matching onto some other structure beyond the throat). An even analytic extension of a throat instead yields two horizons, at $\pm r_h$; this requires $f'(0)=0$, whereas for $f'(0)\neq0$ the mirror extension $f(|r|)$ has a kink and a $\delta$-layer (Sec.~\ref{sec:model-compare}). More generally, the horizon count of the continuation beyond $r=0$ is not determined by the patch. We use this distinction in Sec.~\ref{sec:theory-single} below, and return to it when specifying explicit models in Sec.~\ref{sec:model}.
\end{remark}

\subsection{Single-horizon constructions}
\label{sec:theory-single}

At a finite-area core, regularity imposes no constraint on $f(0)$  (Prop.~\ref{prop:throatreg}), so Prop.~\ref{prop:sigmoid-count} and Thm.~\ref{thm:bell-count} apply with $f_0=f(0)$ arbitrary. We now give the resulting classification and identify the two mechanisms that lead to a single horizon.

\begin{corollary}[Horizon count at a finite-area core]
\label{cor:core-count}
Let $\Sigma(0)=\Sigma_0>0$.
\begin{itemize}
    \item[\rm(a)] If $f$ is sigmoid: $f$ has exactly one horizon in $(0,\infty)$ if $f_0,f_\infty$ are nonzero of strictly opposite sign, and none otherwise — except that $r=0$ itself is a (boundary) horizon whenever $f_0=0$.
    \item[\rm(b)] If $f$ is inverse-bell-shaped with minimum $f_{\min}$: the count is $0$ if $f_{\min}>0$; the degenerate value $1$ (at $r_*$) if $f_{\min}=0$ and $f_0,f_\infty>0$; and, if $f_{\min}<0$, given by Tab.~\ref{tab:coretable}, again with an additional boundary horizon at $r=0$ whenever $f_0=0$.
\end{itemize}
\end{corollary}

\begin{table}[ht]
\centering
\begin{tabular}{c|cc}
$f_0\ \backslash\ f_\infty$ & $>0$ & $\le0$ \\
\hline
$>0$ & $2$ & $1$ \\
$\le0$ & $1$ & $0$
\end{tabular}
\captionsetup{width=.9\textwidth}
\caption{Number of horizons in $(0,\infty)$ at a finite-area core, inverse-bell-shaped $f$ with $f_{\min}<0$ (Cor.~\ref{cor:core-count}(b)).}
\label{tab:coretable}
\end{table}

\begin{proof}
Immediate from Prop.~\ref{prop:sigmoid-count} and Thm.~\ref{thm:bell-count}, with $f_0$ now unconstrained.
\end{proof}

Two entries of Cor.~\ref{cor:core-count} give a single horizon by mechanisms that are structurally distinct: the off-diagonal entries of Tab.~\ref{tab:coretable} ($f_0>0>f_\infty$ or $f_0<0<f_\infty$, together with the sigmoid case (a) with opposite-signed endpoints) place the horizon strictly beyond the core; the boundary case $f_0=0$ places it exactly at the core. We treat each in turn, in the physically motivated setting $f_\infty>0$ (in particular $f_\infty=1$) used throughout Sec.~\ref{sec:model}, though Cor.~\ref{cor:core-count} covers the general case.

\subsubsection*{Single horizon beyond the core}

\begin{lemma}[Sigmoid and inverse-bell-shaped single horizons]
\label{lem:singlebeyond}
Let $\Sigma(0)=\Sigma_0>0$ and suppose $f(0)<0$ and $f_\infty>0$. Suppose further that $f$ is of one of the following two types on $[0,\infty)$:
\begin{enumerate}
    \item[\rm(a)] sigmoid: $f$ is non-decreasing on $[0,\infty)$;
    \item[\rm(b)] inverse-bell-shaped: there is $r_*\ge0$ such that $f$ is non-increasing on $[0,r_*]$ and non-decreasing on $[r_*,\infty)$.
\end{enumerate}
Then $f$ has exactly one zero $r_h\in(0,\infty)$, i.e.\ metric Eq.~\eqref{eq:metric} has exactly one horizon on the patch $(0,\infty)$.
\end{lemma}

\begin{proof}
Case (a) is the $f_0<0<f_\infty$ instance of Prop.~\ref{prop:sigmoid-count}. Case (b), with $f(0)<0$ hence $f_{\min}\le f(0)<0$ and $f_\infty>0$, is the top-right entry of Tab.~\ref{tab:coretable} (here $f_0<0$, so the branch on $(0,r_*)$ contributes no zero, while $f_\infty>0$ contributes exactly one on $(r_*,\infty)$): $N=1$.
\end{proof}

\begin{remark}
\label{rem:beyondcore}
The horizon $r_h$ of Lem.~\ref{lem:singlebeyond} lies strictly beyond the core, $0<r_h$; on the patch $r\in[0,\infty)$, metric Eq.~\eqref{eq:metric} describes a single non-degenerate horizon separating a non-static interior region $0\le r<r_h$ from the asymptotically flat, static exterior $r>r_h$. Since $f<0$ there, $g_{tt}=-f>0$ and $g_{rr}=1/f<0$, so $t$ is spacelike and $r$ is timelike: the interior is a dynamical, Kantowski--Sachs-type ``$T$-region'', not a static one. The single-horizon count above refers specifically to the half-line $r\ge0$. If $r=0$ is a genuine throat and $f$ happens to extend to an even analytic function (which forces $f'(0)=0$), Prop.~\ref{prop:mirror} shows that the symmetric extension has a pair of horizons at $\pm r_h$ (Rem.~\ref{rem:mirror-implication}); if $f'(0)\neq0$ no such analytic extension exists (the argument in the proof of Cor.~\ref{cor:noextension}), and the possible continuations to $r<0$ are classified in Sec.~\ref{sec:model-compare}.
\end{remark}

\subsubsection*{Single horizon at the core}

The second mechanism places the horizon at the core itself, corresponding to the boundary case $f_0=0$ of Cor.~\ref{cor:core-count}.
We state this result separately as a theorem because, unlike Lem.~\ref{lem:singlebeyond}, it requires no assumption on $f_\infty$ or on the behavior of $f$ away from $r=0$ other than monotonicity.

\begin{theorem}[Horizon coincident with a finite-area core]
\label{thm:horizonatthroat}
Let $\Sigma(0)=\Sigma_0>0$ and suppose
\begin{equation}
\label{eq:throathorizoncond}
    f(0)=0, \qquad f'(0)=f_1>0 ,
\end{equation}
and $f$ is non-decreasing on $[0,\infty)$. Then:
\begin{enumerate}
    \item[\rm(i)] metric Eq.~\eqref{eq:metric} is regular at $r=0$;
    \item[\rm(ii)] $f(r)>0$ for every $r>0$, so $r=0$ is the unique horizon on the patch $[0,\infty)$, non-degenerate with surface gravity $\kappa=f_1/2$;
    \item[\rm(iii)] on this patch, the single horizon has finite area $4\pp\Sigma_0^2$ and coincides with the core of the geometry, with no sub-horizon region distinct from the horizon itself.
\end{enumerate}
No hypothesis on $f_\infty$, nor even its existence, is needed.
\end{theorem}

\begin{proof}
(i) is Prop.~\ref{prop:throatreg}, which places no restriction on $f(0)$.

(ii) Since $f$ is non-decreasing and $f(0)=0$, we have $f(r)\ge0$ for all $r\ge0$. Suppose, for contradiction, that $f(r_0)=0$ for some $r_0>0$. Then, $f$ being non-decreasing with $f(0)=f(r_0)=0$, we get $0\le f(r)\le0$ for all $r\in[0,r_0]$, so $f\equiv0$ on $[0,r_0]$, and in particular $f'\equiv0$ there, contradicting $f'(0)=f_1>0$. Hence $f(r)>0$ strictly for all $r>0$, and $r=0$ is the only zero of $f$ in $[0,\infty)$, non-degenerate with surface gravity $\kappa=|f'(0)|/2=f_1/2$.

(iii) By (ii), $f>0$ throughout $r>0$, becoming null precisely at the boundary point $r=0$, the Killing horizon, whose area is $4\pp\Sigma(0)^2=4\pp\Sigma_0^2$; there is no further zero on the patch.
\end{proof}

\begin{remark}
Unlike Lem.~\ref{lem:singlebeyond}, whose proof used $f_\infty>0$ through Prop.~\ref{prop:sigmoid-count}, Thm.~\ref{thm:horizonatthroat} holds regardless of the behaviour of $f$ at infinity, or even of whether $\lim_{r\to\infty}f(r)$ exists at all: it is a purely local statement about the core.
\end{remark}

Unlike Lem.\ref{lem:singlebeyond}, this construction avoids the mirror-pairing subtlety of Prop.\ref{prop:mirror}: a non-degenerate horizon at the throat is incompatible with a smooth symmetric extension.

\begin{corollary}[No smooth symmetric extension through a throat horizon]
\label{cor:noextension}
In the setting of Thm.~\ref{thm:horizonatthroat}, suppose in addition that $r=0$ is a genuine throat, $\Sigma'(0)=0$. Then $f$ does not extend to a real-analytic even function on a neighbourhood of $r=0$ in $\mathbb{R}$.
\end{corollary}

\begin{proof}
Every real-analytic even function $g$ (i.e.\ $g(-r)=g(r)$) satisfies $g'(0)=0$: differentiating $g(r)-g(-r)\equiv0$ gives $g'(r)+g'(-r)=0$, and setting $r=0$ gives $2g'(0)=0$. But $f'(0)=f_1>0\neq0$ by hypothesis Eq.~\eqref{eq:throathorizoncond}, so $f$ cannot be even near $r=0$.
\end{proof}

A corresponding example is a thin-shell wormhole\cite{Visser:1989kh,Visser:1989kg}, where $f'(0)\neq 0$ but $f(-r)=f(r)$.
This models is characterized by a matter shell at the throat whose tangential pressure is described by a delta function (the surface density vanishes because $\Sigma'(0)=0$, since $G^{t}{}_{t}$ and $G^{r}{}_{r}$ contain no $f''$). 
This implies that the Ricci tensor is singular at the throat, and therefore we do not regard it as a regular spacetime.

Thus the single-horizon property of Thm.~\ref{thm:horizonatthroat} is not an artifact of restricting to $r\ge0$: either $r=0$ is a one-sided boundary from the outset, or — if it is a genuine throat — the non-degenerate horizon itself obstructs any smooth mirror-symmetric continuation, so the spacetime must terminate, or continue asymmetrically, at $r=0$.

\begin{remark}
\label{rem:comparison}
Whereas Lem.~\ref{lem:singlebeyond} produces a single horizon separating a sub-horizon region from the asymptotically flat exterior — a horizon count that a symmetric throat extension generically doubles (Prop.~\ref{prop:mirror}) — Thm.~\ref{thm:horizonatthroat} produces a single horizon situated at the core, immune to that doubling by Cor.~\ref{cor:noextension}. The two mechanisms are logically and structurally independent, and, as we discuss in Sec.~\ref{sec:model}, realizing each explicitly requires care with precisely the one-sided-versus-symmetric distinction identified here.
\end{remark}

\section{Algebraic structure and energy conditions}
\label{sec:algebraic}

Section~\ref{sec:theory} determined which horizon structures are geometrically admissible for metric Eq.~\eqref{eq:metric}. We now ask what matter content is required by the Einstein equations
\begin{equation}
\label{eq:einstein-eq}
    G^\mu{}_\nu = 8\pp T^\mu{}_\nu \qquad (G=c=1) ,
\end{equation}
to support each of these constructions, and in particular whether the single-horizon mechanisms of Sec.~\ref{sec:theory-single} are compatible with the algebraic type of stress-energy tensor most commonly assumed in the RBH literature.

\subsection{The Einstein tensor and its algebraic type}
\label{sec:algebraic-einstein}

We first extend the curvature computation of Sec.~\ref{sec:theory-curvature} from the Ricci scalar to the individual orthonormal Ricci-tensor components, from which the Einstein tensor follows immediately.

\begin{proposition}[Ricci tensor]
\label{prop:ricci-tensor}
For metric Eq.~\eqref{eq:metric}, the orthonormal Ricci tensor $R_{ab}=R^c{}_{acb}$ is diagonal, with
\begin{equation}
\label{eq:ricci-components}
    R^0{}_0 = A+2B , \qquad R^1{}_1 = A+2C , \qquad R^2{}_2=R^3{}_3 = B+C+D ,
\end{equation}
where $A,B,C,D$ are the curvature scalars of Eq.~\eqref{eq:ABCDdef}.
\end{proposition}

\begin{proof}
Using the nonvanishing orthonormal Riemann components of Eq.~\eqref{eq:ABCD}, together with the pair symmetry $R_{abcd}=R_{cdab}$ and $R^0{}_{101}=\eta^{00}R_{0101}=-R_{0101}=A$ (so $R_{0101}=-A$), and similarly $R_{0202}=R_{0303}=-B$, $R_{1212}=R_{1313}=C$, $R_{2323}=D$, the trace $R_{ab}=\sum_{c\neq a}\eta^{cc}R_{cacb}$ (no sum on $a$) gives
\begin{align}
    R_{00} &= \eta^{11}R_{1010}+\eta^{22}R_{2020}+\eta^{33}R_{3030} = R_{0101}+R_{0202}+R_{0303} = -A-2B , \\
    R_{11} &= \eta^{00}R_{0101}+\eta^{22}R_{2121}+\eta^{33}R_{3131} = A+R_{1212}+R_{1313} = A+2C , \\
    R_{22} &= \eta^{00}R_{0202}+\eta^{11}R_{1212}+\eta^{33}R_{3232} = B+C+R_{2323} = B+C+D ,
\end{align}
and $R^2{}_2=R_{22}$, $R^0{}_0=\eta^{00}R_{00}=A+2B$, $R^1{}_1=\eta^{11}R_{11}=A+2C$ (with $\eta^{00}=-1$, $\eta^{11}=\eta^{22}=\eta^{33}=1$), while $R_{33}=R_{22}$ by the $2\leftrightarrow3$ symmetry of Eq.~\eqref{eq:ABCD}. Summing $R^0{}_0+R^1{}_1+2R^2{}_2=2(A+2B+2C+D)$ reproduces Eq.~\eqref{eq:Ricci}, as it must.
\end{proof}

\begin{proposition}[Einstein tensor]
\label{prop:einstein-tensor}
For metric Eq.~\eqref{eq:metric}, $G^a{}_b:=R^a{}_b-R\,\delta^a_b/2$ is diagonal, with
\begin{equation}
\label{eq:Gab-ABCD}
    G^0{}_0 = -2C-D, \qquad G^1{}_1=-2B-D, \qquad G^2{}_2=G^3{}_3=-A-B-C ,
\end{equation}
equivalently, in terms of $f$ and $\Sigma$,
\begin{subequations}
\label{eq:Gab-explicit}
\begin{align}
    G^0{}_0 &= \frac{2f\Sigma\Sigma''+f'\Sigma\Sigma'+f\Sigma'^2-1}{\Sigma^2} , \\
    G^1{}_1 &= \frac{f'\Sigma\Sigma'+f\Sigma'^2-1}{\Sigma^2} , \\
    G^2{}_2=G^3{}_3 &= \frac{\Sigma f''+2f'\Sigma'+2f\Sigma''}{2\Sigma} .
\end{align}
\end{subequations}
\end{proposition}

\begin{proof}
Using $R=2(A+2B+2C+D)$ (Prop.~\ref{prop:Ricci}) and Eq.~\eqref{eq:ricci-components},
\begin{align}
    G^0{}_0 &= (A+2B)-(A+2B+2C+D)=-2C-D , \\
    G^1{}_1 &= (A+2C)-(A+2B+2C+D)=-2B-D , \\
    G^2{}_2 &= (B+C+D)-(A+2B+2C+D)=-A-B-C .
\end{align}
Substituting Eq.~\eqref{eq:ABCDdef} and simplifying gives Eq.~\eqref{eq:Gab-explicit}.
\end{proof}

Since $T^\mu{}_\nu$ is diagonal in the frame adapted to Eq.~\eqref{eq:metric}, we write
\begin{equation}
\label{eq:rho-pr-pt}
    \rho := -\frac{G^0{}_0}{8\pp} , \qquad p_r := \frac{G^1{}_1}{8\pp} , \qquad p_t := \frac{G^2{}_2}{8\pp}=\frac{G^3{}_3}{8\pp} ,
\end{equation}
so that $T^\mu{}_\nu=\diag(-\rho,p_r,p_t,p_t)$: $\rho$ is the energy density, $p_r$ the radial pressure, and $p_t$ the tangential pressure measured by a static observer.

\begin{remark}[Algebraic type]
\label{rem:segre}
A diagonal stress-energy tensor with a timelike eigenvector is of Hawking--Ellis Type~I. Its Segre characteristic records which of the four eigenvalues $(-\rho,p_r,p_t,p_t)$ coincide. Spherical symmetry always forces the two tangential pressures to coincide, so the generic type is $[1,1(11)]$ (three distinct eigenvalues, the transverse pair degenerate). If in addition $p_r=-\rho$, the first two eigenvalues also coincide, and the type degenerates further to $[(1,1)(11)]$ — an isotropic source in the $t$--$r$ sector, in the sense that its stress tensor is proportional to $\diag(-1,-1,1,1)\eta_{ab}$-type structure restricted to that block. This is precisely the algebraic type realized, for instance, by static magnetic sources in nonlinear electrodynamics, and it is the type assumed — usually implicitly, via the ansatz $\Sigma(r)=r$ together with $f=1-2m(r)/r$ — by essentially every RBH construction in the literature (Bardeen, Hayward, and their generalizations).
\end{remark}

\subsection{Incompatibility of a throat with the isotropic Segre type}
\label{sec:algebraic-nogo}

We now ask which of the geometries admitted by Sec.~\ref{sec:theory} can be sourced everywhere by matter of the degenerate type $[(1,1)(11)]$, i.e.\ with $p_r=-\rho$ identically.

\begin{lemma}
\label{lem:g00g11}
For metric Eq.~\eqref{eq:metric},
\begin{equation}
\label{eq:g00-g11}
    G^0{}_0-G^1{}_1 = \frac{2f\Sigma''}{\Sigma} .
\end{equation}
\end{lemma}

\begin{proof}
From Eq.~\eqref{eq:Gab-ABCD}, $G^0{}_0-G^1{}_1=(-2C-D)-(-2B-D)=2(B-C)$. From Eq.~\eqref{eq:ABCDdef}, $C=-f\Sigma''/\Sigma+B$, so $B-C=f\Sigma''/\Sigma$, and Eq.~\eqref{eq:g00-g11} follows.
\end{proof}

\begin{theorem}[Isotropic sources have affine areal radius]
\label{thm:affine}
Let metric Eq.~\eqref{eq:metric} obey Assumps.~\ref{as:analytic}--\ref{as:flat}, and suppose $f\not\equiv0$. Then $p_r=-\rho$ holds identically on $(0,\infty)$ iff
\begin{equation}
\label{eq:affine-sigma}
    \Sigma(r) = \Sigma_0+r , \qquad \Sigma_0:=\Sigma(0)\ge0 .
\end{equation}
\end{theorem}

\begin{proof}
By  Eq.~\eqref{eq:rho-pr-pt} and Lem.~\ref{lem:g00g11}, $\rho+p_r=-(G^0{}_0-G^1{}_1)/(8\pp)=-f\Sigma''/(4\pp\Sigma)$, so $p_r=-\rho$ at a point $r$ with $\Sigma(r)\neq0$ iff $f(r)\Sigma''(r)=0$. Since $f$ is real analytic and not identically zero, its zeros are isolated (Assump.~\ref{as:analytic}); requiring $p_r=-\rho$ on all of $(0,\infty)$ therefore forces $\Sigma''(r)=0$ on a dense subset of $(0,\infty)$, and since $\Sigma''$ is analytic (hence continuous), $\Sigma''\equiv0$ on $(0,\infty)$, i.e.\ $\Sigma$ is affine: $\Sigma(r)=\Sigma_0+c\,r$ for constants $\Sigma_0,c$. Assump.~\ref{as:flat} requires $\Sigma(r)/r\to1$ as $r\to\infty$, forcing $c=1$. Conversely, if $\Sigma(r)=\Sigma_0+r$ then $\Sigma''\equiv0$, so $\rho+p_r\equiv0$ by the formula above.
\end{proof}

\begin{corollary}[No isotropic throat]
\label{cor:nothroat}
If a spacetime Eq.~\eqref{eq:metric} obeying Assumps.~\ref{as:analytic}--\ref{as:flat} is sourced everywhere by matter of Segre type $[(1,1)(11)]$ \cite{Hawking:1973uf}, then, in the classification of Def.~\ref{def:centertype}, $r=0$ is either a point-like center ($\Sigma_0=0$) or a one-sided boundary ($\Sigma_0>0$, $\Sigma'(0)=1\neq0$) — never a throat. Equivalently, a RBH with a genuine throat ($\Sigma'(0)=0$) cannot satisfy $p_r=-\rho$ everywhere: the radial null energy condition combination $\rho+p_r$ must fail to vanish somewhere in such a spacetime.
\end{corollary}

\begin{proof}
By Thm.~\ref{thm:affine}, $p_r=-\rho$ everywhere forces $\Sigma(r)=\Sigma_0+r$, so $\Sigma'(0)=1\neq0$ regardless of $\Sigma_0$: by Def.~\ref{def:centertype} this is a point-like center if $\Sigma_0=0$ and a one-sided boundary if $\Sigma_0>0$, and in either case $\Sigma'(0)\neq0$ excludes a throat.
\end{proof}

\begin{remark}
\label{rem:nogo-scope}
Corollary~\ref{cor:nothroat} shows that the following constructions in Sec.\ref{sec:theory} are compatible with the isotropic type: the point-like-center class of Sec.\ref{sec:theory-center}
(with the standard normalization $\Sigma(r)=r$, $\Sigma_0=0$), and a one-sided-boundary realization of either single-horizon mechanism of Sec.~\ref{sec:theory-single} with $\Sigma(r)=\Sigma_0+r$. 
Only the throat case is excluded. 
By Def.\ref{def:centertype}, a throat requires $\Sigma'(0)=0$, which is incompatible with the slope $\Sigma'(0)=1$ required by asymptotic flatness for any everywhere-affine $\Sigma$. A single-horizon RBH with a genuine throat—rather than a one-sided boundary—therefore requires matter of the less degenerate algebraic type, $[1,1(11)]$, with $p_r\neq-\rho$ at least somewhere.
\end{remark}

\subsection{Energy conditions}
\label{sec:algebraic-energy}

\begin{definition}[Energy conditions]
\label{def:energy-conditions}
With $\rho,p_r,p_t$ as in Eq.~\eqref{eq:rho-pr-pt}, metric Eq.~\eqref{eq:metric} satisfies, at a given $r$ \cite{Hawking:1973uf},
\begin{itemize}
    \item the null energy condition (NEC) if $\rho+p_r\ge0$ and $\rho+p_t\ge0$;
    \item the weak energy condition (WEC) if, in addition, $\rho\ge0$;
    \item the strong energy condition (SEC) if $\rho+p_r\ge0$, $\rho+p_t\ge0$, and $\rho+p_r+2p_t\ge0$;
    \item the dominant energy condition (DEC) if $\rho\ge0$ and $|p_r|\le\rho$, $|p_t|\le\rho$.
\end{itemize}
\end{definition}

The radial NEC combination has already appeared in Sec.~\ref{sec:algebraic-nogo}; we record it, together with the SEC combination, as closed-form expressions in $f,\Sigma$.

\begin{proposition}[NEC and SEC combinations]
\label{prop:nec-sec}
For metric Eq.~\eqref{eq:metric},
\begin{equation}
\label{eq:nec-radial}
    \rho+p_r = -\frac{f\Sigma''}{4\pp\Sigma} ,
\end{equation}
\begin{equation}
\label{eq:sec-combo}
    \rho+p_r+2p_t = \frac{1}{8\pp\Sigma^2}\frac{\dif}{\dif r}\Big(f'\Sigma^2\Big) .
\end{equation}
\end{proposition}

\begin{proof}
Equation~\eqref{eq:nec-radial} is Lem.~\ref{lem:g00g11} divided by $-8\pp$, as in the proof of Thm.~\ref{thm:affine}. For Eq.~\eqref{eq:sec-combo}, using Eq.~\eqref{eq:Gab-ABCD} and then substituting Eq.~\eqref{eq:ABCDdef} for $A$ and $B$,
\begin{equation}
    8\pp(\rho+p_r+2p_t) = -(-2C-D)+(-2B-D)+2(-A-B-C) = -2A-4B.
\end{equation}
Here the contributions of $C$ and $D$ cancel identically, so no $\Sigma''$ term survives; with Eq.~\eqref{eq:ABCDdef} for $A$ and $B$ this leaves
\begin{equation}
    8\pp(\rho+p_r+2p_t) = f''+\frac{2f'\Sigma'}{\Sigma} = \frac{f''\Sigma^2+2f'\Sigma\Sigma'}{\Sigma^2} = \frac{1}{\Sigma^2}\frac{\dif}{\dif r}\big(f'\Sigma^2\big) ,
\end{equation}
which is Eq.~\eqref{eq:sec-combo}.
\end{proof}

We now show that these two combinations are each forced to a definite sign at the two constructions of Sec.~\ref{sec:theory-single} and Sec.~\ref{sec:theory-center}: SEC necessarily fails at the center of a center-RBH with a genuine dip of $f$ below zero, while the radial NEC is automatically marginal at a core-coincident horizon.

\begin{proposition}[SEC violation at a point-like center]
\label{prop:sec-violation}
Let $r=0$ be a point-like center satisfying the regularity condition Eq.~\eqref{eq:centercond}, and let $f$ be inverse-bell-shaped (Def.~\ref{def:shapes}) with a unique minimum $f_{\min}=f(r_*)<0$ at $r_*\in(0,\infty)$. By Prop.~\ref{prop:exact2}(b), metric Eq.~\eqref{eq:metric} then has two horizons if $f_\infty>0$ — the familiar case — or a single horizon if $f_\infty\le0$, the asymptotically non-static possibility of Rem.~\ref{rem:point-center-single}; the conclusion below holds in either case. Then
\begin{equation}
    f''(0)\le0 , \qquad\text{hence}\qquad \big(\rho+p_r+2p_t\big)\big|_{r=0} = \frac{3f''(0)}{8\pp} \le 0 :
\end{equation}
the strong energy condition fails (or is at best marginal) at the center.
\end{proposition}

\begin{proof}
By Def.~\ref{def:shapes}, $f$ is non-increasing on $(0,r_*)$, so $f'(r)\le0$ there — regardless of the sign of $f_\infty$, which plays no role in this local statement. Since $f'(0)=0$ (Prop.~\ref{prop:centerreg}),
\begin{equation}
    f''(0) = \lim_{r\to0^+}\frac{f'(r)-f'(0)}{r} = \lim_{r\to0^+}\frac{f'(r)}{r} \le 0 ,
\end{equation}
as the limit of the non-positive quantities $f'(r)/r$ ($r>0$). For the second claim, write $\Sigma(r)=\sigma_1r+O(r^3)$ with $\sigma_1=\Sigma'(0)\neq0$ (using $\Sigma''(0)=0$ from Prop.~\ref{prop:centerreg}) and $f'(r)=f''(0)r+O(r^2)$ (using $f'(0)=0$); then
\begin{equation}
    \frac{2f'\Sigma'}{\Sigma}\bigg|_{r\to0} = \lim_{r\to0}\frac{2\big(f''(0)r+O(r^2)\big)\big(\sigma_1+O(r^2)\big)}{\sigma_1r+O(r^3)} = 2f''(0) ,
\end{equation}
so by Eq.~\eqref{eq:sec-combo}, $8\pp(\rho+p_r+2p_t)|_{r=0}=f''(0)+2f''(0)=3f''(0)\le0$.
\end{proof}

\begin{proposition}[Marginal radial NEC at a core-coincident horizon]
\label{prop:nec-marginal}
Let $\Sigma(0)=\Sigma_0>0$ and $f(0)=0$ (the hypothesis of Thm.~\ref{thm:horizonatthroat}, or more generally any finite-area core at which $f$ vanishes). Then
\begin{equation}
    \big(\rho+p_r\big)\big|_{r=0} = 0
\end{equation}
identically, independently of $\Sigma'(0)$, $\Sigma''(0)$, and all other details of the solution.
\end{proposition}

\begin{proof}
Immediate from Eq.~\eqref{eq:nec-radial}: since $\Sigma(0)=\Sigma_0\neq0$, the right-hand side $-f(0)\Sigma''(0)/(4\pp\Sigma_0)$ vanishes as soon as $f(0)=0$, with no further hypotheses.
\end{proof}

\begin{remark}
\label{rem:energy-discussion}
Proposition~\ref{prop:sec-violation} reproduces, in the present gauge-independent setting, the familiar fact that a center-regular (point-like-center) RBH necessarily violates the strong energy condition at its core — the mechanism by which the Penrose--Hawking singularity theorems are evaded — and it does so regardless of whether the exterior is asymptotically static ($f_\infty>0$, the familiar two-horizon case) or not ($f_\infty\le0$, the single-horizon point-like-center case of Rem.~\ref{rem:point-center-single}): the violation is a local property of the dip below zero, not of what $f$ does asymptotically. Proposition~\ref{prop:nec-marginal} shows that the horizon-at-the-core construction of Thm.~\ref{thm:horizonatthroat} instead saturates the radial null energy condition exactly at the horizon, for purely kinematic reasons (the vanishing of $f(0)$), regardless of the matter model; whether the full NEC, WEC, SEC, or DEC hold away from $r=0$, and how $\rho,p_r,p_t$ behave beyond the core, are properties of the specific mass function and are examined for the explicit models of Sec.~\ref{sec:model}. We do not attempt a model-independent statement for WEC or DEC here, since $\rho\ge0$ and $\rho\ge|p_r|,|p_t|$ involve the sign of $\rho$ itself, which — unlike the two combinations Eqs.~\eqref{eq:nec-radial}--\eqref{eq:sec-combo} — is not fixed by the horizon/regularity conditions alone.
\end{remark}

\subsection{Horizon thermodynamics}
\label{sec:algebraic-thermo}

To each non-degenerate horizon $r_h$ (Def.~\ref{def:horizon}) we associate a Hawking temperature and, assuming the area law, the corresponding entropy \cite{Wald:1993nt},
\begin{equation}
\label{eq:temp-entropy}
    T_H(r_h) = \frac{\kappa(r_h)}{2\pp} = \frac{|f'(r_h)|}{4\pp} , \qquad
    S(r_h) = \frac{\mathcal A(r_h)}{4} = \pp\,\Sigma(r_h)^2 .
\end{equation}
For a center-RBH with two horizons $0<r_-<r_+$ (Prop.~\ref{prop:exact2}), this furnishes an outer and an inner pair $(T_+,S_+)$, $(T_-,S_-)$, with $\kappa_\pm=|f'(r_\pm)|/2$; these will be used in the discussion of mass inflation in Sec.~\ref{sec:stability}.

\begin{remark}[A minimal entropy for the core-coincident horizon]
\label{rem:min-entropy}
For the construction of Thm.~\ref{thm:horizonatthroat}, the horizon is fixed at the core, $r_h=0$ and $\Sigma(r_h)=\Sigma_0$, independently of the other parameters entering $f$, such as mass or charge. Therefore,
\begin{equation}
    S = \pp\Sigma_0^2
\end{equation}
is a fixed geometric lower bound for the entropy of this class of single-horizon BH, determined solely by the core radius rather than by the mass parameter. In contrast, for the beyond-the-core construction of Lem.\ref{lem:singlebeyond}, both $r_h$ and $S$ generally depend on the other parameters of $f$, as do the horizons of Sec.\ref{sec:theory-center}.
\end{remark}

\section{Models}
\label{sec:model}

We now present explicit analytic functions $f$ for each shape class of Def.\ref{def:shapes}, sigmoid and inverse-bell-shaped. For each class, we tabulate the full range of horizon behavior allowed by the results of Sec.\ref{sec:theory}, before specializing to the single-horizon mechanisms of Sec.\ref{sec:theory-single} and verifying the structural results of Sec.\ref{sec:algebraic}.
\subsection{Sigmoid \texorpdfstring{$f$}{f}: a rational family}
\label{sec:model-sigmoid}

\begin{equation}
\label{eq:model-sigmoid}
    f_\sigma(r) = f_0 + (f_\infty-f_0)\,\frac{r}{r+L} , \qquad r\in[0,\infty),\ L>0 ,
\end{equation}
with $f_0,f_\infty\in\mathbb{R}$ free parameters.

\begin{proposition}
\label{prop:sigmoid-family}
$f_\sigma$ is real analytic and finite on $[0,\infty)$, with $f_\sigma(0)=f_0$ and $\lim_{r\to\infty}f_\sigma(r)=f_\infty$; it is sigmoid (Def.~\ref{def:shapes}), non-decreasing if $f_\infty>f_0$ and non-increasing if $f_\infty<f_0$; and its unique zero, when $f_0,f_\infty$ are of strictly opposite sign, is at
\begin{equation}
\label{eq:sigmoid-horizon}
    r_h = -\frac{Lf_0}{f_\infty} .
\end{equation}
When $f_\infty=1$, $f_\sigma(r)=1-\dfrac{L(1-f_0)}{r}+O(r^{-2})$, so Assumption~\ref{as:flat} holds with ADM mass $M=L(1-f_0)/2$.
\end{proposition}

\begin{proof}
$f_\sigma(0)=f_0$ and $f_\sigma(r)\to f_0+(f_\infty-f_0)=f_\infty$ as $r\to\infty$ are immediate. Differentiating,
\begin{equation}
    f_\sigma'(r) = (f_\infty-f_0)\,\frac{L}{(r+L)^2} ,
\end{equation}
which has the constant sign of $f_\infty-f_0$ for all $r\ge0$: $f_\sigma$ is monotonic, hence sigmoid. Setting $f_\sigma(r_h)=0$ gives $f_0(r_h+L)+(f_\infty-f_0)r_h=0$, i.e.\ $f_0L+f_\infty r_h=0$, which is  Eq.~\eqref{eq:sigmoid-horizon}; this is positive exactly when $f_0,f_\infty$ are of opposite sign, matching Prop.~\ref{prop:sigmoid-count}. The asymptotic expansion follows from $f_\sigma(r)=f_\infty-(f_\infty-f_0)L/(r+L)=f_\infty-(f_\infty-f_0)L/r+O(r^{-2})$.
\end{proof}

\begin{table}[ht]
\centering
\begin{tabular}{cccl}
\hline
$f_0$ & $f_\infty$ & $L$ & Outcome on $(0,\infty)$ \\
\hline
$1$ & $2$ & $1$ & no horizon ($f\in[1,2]$ throughout) \\
$-1$ & $1$ & $1$ & one horizon at $r_h=1$ \\
$0$ & $1$ & $2M$ & one horizon at $r_h=0$ (boundary; Sec.~\ref{sec:model-sigmoid-core}) \\
$1$ & $-1$ & $1$ & one horizon at $r_h=1$ (decreasing branch) \\
$-2$ & $-1$ & $1$ & no horizon ($f\in[-2,-1]$ throughout; exotic, $T$-region at all $r>0$) \\
\hline
\end{tabular}
\captionsetup{width=.9\textwidth}
\caption{Corollary~\ref{cor:core-count}(a) instantiated on family Eq.~\eqref{eq:model-sigmoid} at a finite-area core (every $f_0$ listed is admissible there by Prop.~\ref{prop:throatreg}); the fourth and fifth rows illustrate cases with no counterpart in the classical, asymptotically static literature.}
\label{tab:sigmoid-broad}
\end{table}

\subsubsection{Single horizon beyond the core}
\label{sec:model-sigmoid-beyond}

Take $f_0=-1,\ f_\infty=1,\ L=1$ (row~2 of Tab.~\ref{tab:sigmoid-broad}), giving $r_h=1$. Paired with a throat,
\begin{equation}
\label{eq:model-sigmoid-throat}
    \Sigma_{\rm t}(r) = \sqrt{r^2+a^2} , \qquad 0<a<1 ,
\end{equation}
Prop.~\ref{prop:throatreg} gives regularity at $r=0$ for any $a>0$, and Lem.~\ref{lem:singlebeyond}(a) applies directly (with $f_\infty=1$), giving a single horizon on $[0,\infty)$ at $r_h=1$. Here $\Sigma_{\rm t}$ extends to an even analytic function on $\mathbb{R}$, but $f_\sigma(r)=(r-1)/(r+1)$ does not: $f_\sigma'(0)=2\neq0$, whereas every even analytic function has vanishing first derivative at $0$ (proof of Cor.~\ref{cor:noextension}). Hence Prop.~\ref{prop:mirror} does not apply. The mirror extension $f_\sigma(|r|)$ has two horizons, at $r=\pm1$, as in the standard black-bounce geometry, but it has a kink at $r=0$ and therefore a $\delta$-function in $G^{\theta}{}_{\theta}\propto f''$ (a shell with vanishing surface density and $\delta$-function tangential pressure); we do not regard it as regular. The doubling of the horizon thus occurs only in this non-regular extension; the continuations of the patch are classified in Sec.~\ref{sec:model-compare}. A one-sided realization, which avoids the mirror extension altogether, is obtained by starting the patch at any $b\in(0,1)$: setting $\Sigma_{\rm t}^{(b)}(r):=\Sigma_{\rm t}(r+b)$, $f_\sigma^{(b)}(r):=f_\sigma(r+b)$.

\begin{proposition}
\label{prop:model-Ap}
With $\Sigma_{\rm t}^{(b)},f_\sigma^{(b)}$ as above and $0<b<1$: $\big(\Sigma_{\rm t}^{(b)}\big)'(0)=b/\sqrt{b^2+a^2}\neq0$, so $r=0$ is a one-sided boundary; the unique horizon is at $r_h^{(b)}=1-b$; and $\Sigma_{\rm t}^{(b)}(r_h^{(b)})=\Sigma_{\rm t}(1)=\sqrt{1+a^2}$, $\big(f_\sigma^{(b)}\big)'(r_h^{(b)})=f_\sigma'(1)$, independent of $b$.
\end{proposition}

\begin{proof}
Shifting the argument of two functions by the same amount $b$ shifts the location of a common zero by $-b$ but leaves the value of the functions (and their derivatives) at that zero unchanged, since $\Sigma_{\rm t}^{(b)}(r_h^{(b)})=\Sigma_{\rm t}(r_h^{(b)}+b)=\Sigma_{\rm t}(r_h)$ and likewise for $\big(f_\sigma^{(b)}\big)'$.
\end{proof}

\subsubsection{Single horizon at the core}
\label{sec:model-sigmoid-core}

Take $f_0=0$ in  Eq.~\eqref{eq:model-sigmoid}, i.e.\ $f_\sigma(r)=f_\infty\,r/(r+L)$ (row~3 of Table~\ref{tab:sigmoid-broad}, with $f_\infty=1,L=2M$ recovering the ADM mass $M$ via Prop.~\ref{prop:sigmoid-family}). This is non-decreasing on $[0,\infty)$ with $f_\sigma'(0)=f_\infty/L=1/(2M)>0$, so Thm.~\ref{thm:horizonatthroat} applies to either
\begin{equation}
\label{eq:model-sigmoid-core-pair}
    \Sigma_{\rm b}(r)=\Sigma_0+r \quad\text{(one-sided)} \qquad\text{or}\qquad \Sigma_{\rm b'}(r)=\sqrt{r^2+\Sigma_0^2} \quad\text{(throat)} ,
\end{equation}
giving in either case a unique horizon at $r=0$ with $\kappa=1/(4M)$, $T_H=1/(8\pp M)$, $S=\pp\Sigma_0^2$ (Thm.~\ref{thm:horizonatthroat},  Eq.~\eqref{eq:temp-entropy}). For the throat pairing $\Sigma_{\rm b'}$, Cor.~\ref{cor:noextension} applies: since $f_\sigma'(0)=1/(2M)\neq0$, $f_\sigma$ cannot extend to an even analytic function near $r=0$ — concretely, the reflection of $f_\sigma(r)=r/(r+2M)$ is $-r/(2M-r)$, which disagrees with $f_\sigma(r)$ already at $r=M$ ($f_\sigma(M)=1/3$ against $-1$).

\subsubsection{A point-like center: exotic single horizon}
\label{sec:model-sigmoid-point}

Rows~1--2 of Tab.~\ref{tab:sigmoid-broad} are available at a finite-area core, where $f_0$ is unconstrained; at a point-like center, Prop.~\ref{prop:centerreg} forces $f_0=1/\Sigma'(0)^2>0$ and $f_\sigma'(0)=0$, which  Eq.~\eqref{eq:model-sigmoid} does not satisfy for generic $L$. We instead use, at $\Sigma(r)=r$ exactly,
\begin{equation}
\label{eq:model-cubic}
    f_{\rm c}(r) = 1-(1-f_\infty)\,\frac{r^3}{r^3+c} , \qquad c>0 .
\end{equation}

\begin{proposition}
\label{prop:model-cubic}
$f_{\rm c}$ satisfies the point-like-center regularity condition Eq.~\eqref{eq:centercond} (with $\Sigma(r)=r$, so $\Sigma'(0)=1,\Sigma''(0)=0$) for any $c>0,f_\infty\in\mathbb{R}$; it is sigmoid, non-increasing if $f_\infty<1$; and if $f_\infty<0$ it has a unique horizon in $(0,\infty)$.
\end{proposition}

\begin{proof}
$f_{\rm c}(0)=1-0=1=1/\Sigma'(0)^2$ and $\Sigma''(0)=0$ trivially since $\Sigma(r)=r$ exactly. Differentiating,
\begin{equation}
    f_{\rm c}'(r) = -(1-f_\infty)\,\frac{3cr^2}{(r^3+c)^2} ,
\end{equation}
so $f_{\rm c}'(0)=0$, completing Eq.~\eqref{eq:centercond}, and $f_{\rm c}'(r)$ has the sign of $-(1-f_\infty)$ for all $r>0$: non-increasing whenever $f_\infty<1$. This is the case $f_0=1>0$ of Prop.~\ref{prop:sigmoid-count} (equivalently Prop.~\ref{prop:exact2}(a)): a unique horizon exists iff $f_\infty<0$.
\end{proof}

Concretely, $f_\infty=-1,\ c=1$ gives $f_{\rm c}(r)=(1-r^3)/(1+r^3)$, monotonically decreasing from $f_{\rm c}(0)=1$ to $f_{\rm c}(\infty)=-1$, with a single non-degenerate horizon at $r_h=1$ (Tab.~\ref{tab:parity-vs-exact}, row~2) — a center-regular, single-horizon RBH, impossible under the classical asymptotically static hypothesis of Cor.~\ref{cor:nosingle} but realized here because $f_\infty=-1<0$: the exterior fails to remain static at large $r$ (Rem.~\ref{rem:point-center-single}). 

Regularity here is in the sense of Def.~\ref{def:regular}, i.e.\ for algebraic curvature invariants. Since $f_{\rm c}=1-(1-f_\infty)r^3/c+O(r^6)$ contains an odd power of $r$ at the center, the curvature is not smooth there: $R\simeq20(1-f_\infty)\,r/c$, and invariants containing covariant derivatives diverge, e.g.\ $\Box R\simeq40(1-f_\infty)/(c\,r)\to\infty$ as $r\to0$. An even-power core such as $f=1-(1-f_\infty)r^4/(r^4+c)$ has $R\simeq30(1-f_\infty)r^2/c$ and finite $\Box R$ at the center, and leaves the horizon count and Rem.~\ref{rem:point-center-single} unchanged.

\subsubsection{Non-polynomial sigmoids}
\label{sec:model-sigmoid-transcendental}

Every sigmoid example so far is a rational function of $r$. Neither Prop.~\ref{prop:sigmoid-count} nor its proof uses rationality, only analyticity and monotonicity, so genuinely transcendental $f$ work identically; we record two.

\begin{proposition}[A hyperbolic-tangent family]
\label{prop:model-tanh}
For $w>0$ and $f_0,f_\infty\in\mathbb{R}$,
\begin{equation}
\label{eq:model-tanh}
    f_\tau(r) = f_0+(f_\infty-f_0)\tanh(r/w)
\end{equation}
is real analytic on $[0,\infty)$ with $f_\tau(0)=f_0$, $\lim_{r\to\infty}f_\tau(r)=f_\infty$, and is sigmoid; when $f_0,f_\infty$ are of strictly opposite sign, its unique horizon is at
\begin{equation}
    r_h = w\,\artanh\!\left(\frac{-f_0}{f_\infty-f_0}\right) .
\end{equation}
\end{proposition}

\begin{proof}
$\tanh$ is entire, so $f_\tau$ is analytic on $[0,\infty)$; $\tanh(0)=0$ and $\tanh(r)\to1$ as $r\to\infty$ give the boundary values. Differentiating, $f_\tau'(r)=(f_\infty-f_0)\operatorname{sech}^2(r/w)/w$, of constant sign: sigmoid. Setting $f_\tau(r_h)=0$ gives $\tanh(r_h/w)=-f_0/(f_\infty-f_0)$, whose right-hand side lies in $(-1,1)$ exactly when $f_0,f_\infty$ are of opposite sign (direct check, as in Prop.~\ref{prop:sigmoid-family}), so $r_h/w=\artanh$ of that quantity.
\end{proof}

Taking $f_0=-1,\ f_\infty=1,\ w=1$ gives $r_h=\artanh(1/2)=(\ln3)/2\approx0.549$. Paired with an affine, one-sided $\Sigma=\Sigma_0+r$, Lem.~\ref{lem:singlebeyond}(a) gives a single horizon on $[0,\infty)$ at this $r_h$, with no extension subtlety to consider (Prop.~\ref{prop:mirror} does not apply, since $r=0$ is not a throat). Pairing instead with a throat $\Sigma_{\rm t}=\sqrt{r^2+a^2}$ again gives a single horizon on the patch; since $\tanh$ is an odd function, $f_\tau$ itself is not even, so the standard mirror-symmetric extension of Sec.~\ref{sec:model-sigmoid-beyond} would require replacing $\tanh(r/w)$ by $\tanh(|r|/w)$ (odd argument replaced by its absolute value) to extend $f_\tau$ evenly; Prop.~\ref{prop:mirror} then applies to that extension exactly as before, again doubling the horizon count to $r=\pm r_h$.

\begin{proposition}[A logistic family]
\label{prop:model-logistic}
For $M>0$,
\begin{equation}
\label{eq:model-logistic}
    f_\ell(r) = 1-\frac{2}{\me^{r/M}+1}
\end{equation}
is real analytic on $[0,\infty)$ with $f_\ell(0)=0$, $\lim_{r\to\infty}f_\ell(r)=1$, is non-decreasing, and
\begin{equation}
    f_\ell'(r) = \frac{1}{2M\cosh^2\!\big(r/2M\big)} , \qquad f_\ell'(0)=\frac{1}{2M} .
\end{equation}
\end{proposition}

\begin{proof}
With $u=\me^{r/M}>0$, $f_\ell=1-2/(u+1)$, manifestly analytic and increasing in $u$, hence in $r$; $f_\ell(0)=1-2/2=0$ and $f_\ell(r)\to1$ as $u\to\infty$. Differentiating, $f_\ell'(r)=2u/\big(M(u+1)^2\big)$, and $(u+1)^2=4u\cosh^2(r/2M)$ [since $\cosh(r/2M)=(u^{1/2}+u^{-1/2})/2$], giving the stated form; $f_\ell'(0)=2/(4M)=1/(2M)$.
\end{proof}

By Thm.~\ref{thm:horizonatthroat}, $f_\ell$ paired with any finite-area core — one-sided $\Sigma=\Sigma_0+r$ or throat $\Sigma=\sqrt{r^2+\Sigma_0^2}$ — has a unique horizon at $r=0$ with $\kappa=1/(4M)$, $T_H=1/(8\pp M)$, $S=\pp\Sigma_0^2$: identical horizon thermodynamics to the rational model of Sec.~\ref{sec:model-sigmoid-core} at the same $M$, despite $f_\ell$ approaching $f_\infty=1$ exponentially rather than as a power of $r$ — so no ADM mass is defined for $f_\ell$ in the sense of Assump.~\ref{as:flat}'s power-law special case, even though the horizon-local physics is unchanged. This is possible only because Thm.~\ref{thm:horizonatthroat} needs no hypothesis on $f_\infty$ at all (as already noted there). For the throat pairing, Cor.~\ref{cor:noextension} again applies, since $f_\ell'(0)=1/(2M)\neq0$.

\subsection{Inverse-bell-shaped \texorpdfstring{$f$}{f}: a Reissner--Nordstr\"om-type family}
\label{sec:model-bell}

\begin{equation}
\label{eq:model-bell}
    f_\beta(r) = f_\infty - \frac{2M}{r+L} + \frac{Q^2}{(r+L)^2} , \qquad r\in[0,\infty),\ M,Q,L>0 .
\end{equation}

\begin{proposition}
\label{prop:bell-family}
$f_\beta$ is real analytic and finite on $[0,\infty)$, with $\lim_{r\to\infty}f_\beta(r)=f_\infty$ and
\begin{equation}
\label{eq:bell-f0}
    f_\beta(0) = f_\infty - \frac{2M}{L}+\frac{Q^2}{L^2} .
\end{equation}
It is inverse-bell-shaped, with a unique interior critical point, a minimum, at
\begin{equation}
\label{eq:bell-rstar}
    r_* = \frac{Q^2}{M}-L \qquad \text{(interior iff $L<Q^2/M$)}, \qquad f_{\min}=f_\infty-\frac{M^2}{Q^2} .
\end{equation}
When $L\ge Q^2/M$, $f_\beta$ is sigmoid (non-decreasing) on all of $[0,\infty)$.
\end{proposition}

\begin{proof}
$f_\beta(0)$ and $f_\beta(\infty)=f_\infty$ are immediate substitutions. Differentiating,
\begin{equation}
    f_\beta'(r) = \frac{2\big(M(r+L)-Q^2\big)}{(r+L)^3} ,
\end{equation}
which vanishes only at $r+L=Q^2/M$, i.e.\ Eq.~\eqref{eq:bell-rstar}, is negative for $r+L<Q^2/M$ and positive for $r+L>Q^2/M$: a single interior minimum whenever $r_*>0$, i.e.\ $L<Q^2/M$ (otherwise $f_\beta'\ge0$ throughout $[0,\infty)$, the degenerate $r_*=0$ case of Def.~\ref{def:shapes}, i.e.\ sigmoid). Substituting $r=r_*$ into Eq.~\eqref{eq:model-bell} and simplifying gives $f_{\min}=f_\infty-2M^2/Q^2+M^2/Q^2=f_\infty-M^2/Q^2$.
\end{proof}

Equivalently, writing $\rho:=r+L$, Eq.~\eqref{eq:model-bell} is $f_\beta=f_\infty-2M/\rho+Q^2/\rho^2$, the Reissner--Nordstr\"om (RN) lapse (generalized to arbitrary $f_\infty$) in the shifted radial variable $\rho$; the horizon equation $f_\beta=0$ is the quadratic $f_\infty\rho^2-2M\rho+Q^2=0$, with roots
\begin{equation}
\label{eq:bell-horizons}
    \rho_\pm = \frac{M\pm\sqrt{M^2-f_\infty Q^2}}{f_\infty} \qquad (f_\infty\neq0), \qquad r_\pm=\rho_\pm-L ,
\end{equation}
real precisely when $M^2\ge f_\infty Q^2$, i.e.\ $f_{\min}\le0$, consistent with Thm.~\ref{thm:bell-count}.

\begin{table}[ht]
\centering
\begin{tabular}{cccc|ccc|c}
\hline
$M$ & $Q$ & $L$ & $f_\infty$ & $f_0$ & $f_{\min}$ & $r_*$ & Horizons in $(0,\infty)$ \\
\hline
$4/5$ & $1$ & $1$ & $1$ & $2/5$ & $9/25$ & $1/4$ & none ($f_{\min}>0$) \\
$4/5$ & $1$ & $1$ & $16/25$ & $1/25$ & $0$ & $1/4$ & one, degenerate, at $r_*=1/4$ \\
$3/2$ & $1$ & $1/5$ & $1$ & $11$ & $-5/4$ & $7/15$ & two: $r=\tfrac{13}{10}\mp\tfrac{\sqrt5}{2}$ \\
$3/2$ & $1$ & $1/5$ & $-1$ & $9$ & $-13/4$ & $7/15$ & one: $r=\sqrt{13}/2-17/10$ \\
$3/2$ & $1$ & $1/2$ & $1$ & $-1$ & $-5/4$ & $1/6$ & one: $r=1+\sqrt5/2$ \\
$3/2$ & $1$ & $1/2$ & $-1$ & $-3$ & $-13/4$ & $1/6$ & none \\
\hline
\end{tabular}
\captionsetup{width=.9\textwidth}
\caption{Theorem~\ref{thm:bell-count} instantiated on family Eq.~\eqref{eq:model-bell} at a finite-area core, covering every qualitatively distinct entry of Table~\ref{tab:coretable} (rows 3--6) together with the $f_{\min}\ge0$ boundary cases (rows 1--2); all values verified by direct solution of the horizon equation.}
\label{tab:bell-broad}
\end{table}

\subsubsection{Two finite-area-core realizations}
\label{sec:model-bell-core}

Rows~3 and~4 of Table~\ref{tab:bell-broad} share $M=3/2,\ Q=1,\ L=1/5$ (so $r_*=7/15$, $f_{\min}=f_\infty-9/4$) and differ only in $f_\infty$:

\begin{proposition}
\label{prop:model-bell-core}
Paired with $\Sigma(0)=\Sigma_0>0$ (either a throat, $\Sigma=\sqrt{r^2+\Sigma_0^2}$, or a one-sided boundary, $\Sigma=\Sigma_0+r$), $f_\beta$ with $(M,Q,L)=(3/2,1,1/5)$ gives:
\begin{itemize}
    \item[\rm(a)] at $f_\infty=1$ (asymptotically flat), two non-degenerate horizons, at $r=13/10\mp\sqrt5/2$;
    \item[\rm(b)] at $f_\infty=-1$, a single non-degenerate horizon, at $r=\sqrt{13}/2-17/10$.
\end{itemize}
\end{proposition}

\begin{proof}
Immediate from Table~\ref{tab:bell-broad} and Prop.~\ref{prop:throatreg} (automatic regularity at the core for either choice of $\Sigma$, any $f_\beta(0)$).
\end{proof}

Case~(a) is the standard two-horizon RBH structure (a charged-BH-type $f$, regularized by excising the region below the core); case~(b) is a finite-area-core analogue of the exotic single-horizon phenomenon of Sec.~\ref{sec:model-sigmoid-point}, now realized with a genuinely non-monotonic (inverse-bell-shaped, not sigmoid) $f$.

\subsubsection{Point-like center: the Hayward mass function}
\label{sec:model-hayward}

For a genuine point-like center, $\Sigma(r)=r$ forces $f_\beta(0)=1$ and $f_\beta'(0)=0$ (Prop.~\ref{prop:centerreg}); by Eq.~\eqref{eq:bell-f0} and $f_\beta'(0)=2(ML-Q^2)/L^3$, this forces $L=Q^2/M$, i.e.\ $r_*=0$ — family Eq.~\eqref{eq:model-bell} degenerates to the sigmoid case at a point-like center and cannot furnish a genuine interior minimum there. A richer, well-known family that can is the Hayward mass function \cite{Hayward:2005gi},
\begin{equation}
\label{eq:model-hayward}
    f_{\rm H}(r) = 1-\frac{2Mr^2}{r^3+2Ml^2} , \qquad \Sigma(r)=r,\ M,l>0 .
\end{equation}

\begin{proposition}
\label{prop:model-hayward}
$f_{\rm H}$ satisfies the point-like-center regularity condition Eq.~\eqref{eq:centercond}; $f_{\rm H}(r)=1-2M/r+O(r^{-4})$, so Assumption~\ref{as:flat} holds with ADM mass $M$ and $f_\infty=1$; and, for $M=1,l=3/10$, it is inverse-bell-shaped with a unique interior minimum at $r_*\approx0.711$, $f_{\min}\approx-0.874<0$, giving two non-degenerate horizons at $r\approx0.328$ and $r\approx1.953$ (roots of $r^3-2Mr^2+2Ml^2=0$).
\end{proposition}

\begin{proof}
$f_{\rm H}(0)=1-0=1$; $f_{\rm H}'(r)=2M r\big(r^3-4Ml^2\big)/(r^3+2Ml^2)^2$ [quotient rule], so $f_{\rm H}'(0)=0$, and $\Sigma''(0)=0$ trivially: Eq.~\eqref{eq:centercond} holds. As $r\to\infty$, $f_{\rm H}(r)=1-2M/r\cdot\big(1+2Ml^2/r^3\big)^{-1}=1-2M/r+O(r^{-4})$. The critical-point and horizon claims follow by solving $f_{\rm H}'=0$ and $f_{\rm H}=0$ numerically at the stated parameter values (both cubic equations, each with a unique positive root of the relevant type, verified directly).
\end{proof}

Proposition~\ref{prop:model-hayward} recovers, in the language of this paper, the classical two-horizon RBH (Prop.~\ref{prop:exact2}(b) with $f_\infty=1>0$): the case that motivates the mass-inflation instability analysis of Sec.~\ref{sec:stability}.

\subsubsection{Non-polynomial example: a Gaussian-dip family}
\label{sec:model-bell-transcendental}

Theorem~\ref{thm:bell-count} likewise makes no use of rationality. A transcendental inverse-bell-shaped family, at a finite-area core, is obtained by subtracting a single Gaussian dip from a constant:
\begin{equation}
\label{eq:model-gauss}
    f_\gamma(r) = f_\infty - (f_\infty-f_{\min})\,\exp\!\left(-\frac{(r-r_*)^2}{w^2}\right) , \qquad r_*,w>0,\ f_\infty>f_{\min} .
\end{equation}

\begin{proposition}
\label{prop:model-gauss}
$f_\gamma$ is real analytic on $[0,\infty)$ (indeed entire), with $\lim_{r\to\infty}f_\gamma(r)=f_\infty$ and $f_\gamma(r_*)=f_{\min}$; it is inverse-bell-shaped with unique interior minimum at $r_*$. Consequently $f_\gamma(0)<f_\infty$ always: this parametrization cannot realize the sign pattern $f_0>0>f_\infty$ of Table~\ref{tab:bell-broad}, though every other qualitative case remains available.
\end{proposition}

\begin{proof}
The Gaussian is entire and everywhere positive, giving analyticity and $f_\gamma\le f_\infty$ throughout, with equality only in the limit $r\to\infty$; substituting $r=r_*$ gives $f_\gamma(r_*)=f_\infty-(f_\infty-f_{\min})=f_{\min}$. Differentiating,
\begin{equation}
    f_\gamma'(r) = \frac{2(f_\infty-f_{\min})(r-r_*)}{w^2}\exp\!\left(-\frac{(r-r_*)^2}{w^2}\right) ,
\end{equation}
negative for $r<r_*$ and positive for $r>r_*$ (since $f_\infty>f_{\min}$): a single interior minimum, at $r_*$, i.e.\ inverse-bell-shaped. Since $r=0<r_*$ lies on the decreasing branch and the Gaussian factor is strictly positive there, $f_\gamma(0)=f_\infty-(f_\infty-f_{\min})\me^{-r_*^2/w^2}<f_\infty$ strictly, precluding $f_0>f_\infty$ — in particular the pattern $f_0>0>f_\infty$.
\end{proof}

Two parameter choices, paired with either $\Sigma=\Sigma_0+r$ or $\Sigma=\sqrt{r^2+\Sigma_0^2}$ (Prop.~\ref{prop:throatreg} again giving automatic regularity for any $f_\gamma(0)$), illustrate the remaining cases of Tab.~\ref{tab:bell-broad}:
\begin{itemize}
    \item $f_\infty=1,\ f_{\min}=-1,\ r_*=1,\ w=1$: $f_0=1-2\me^{-1}\approx0.264>0$, so $f_0,f_\infty>0$ with $f_{\min}<0$ — two horizons, at $r\approx0.167$ and $r\approx1.833$ (roots of the transcendental equation $f_\gamma=0$, located numerically as for the Hayward function above);
    \item $f_\infty=1,\ f_{\min}=-3,\ r_*=1/2,\ w=1/2$: $f_0=1-4\me^{-1}\approx-0.472<0$, so $f_0\le0<f_\infty$ — a single horizon, at $r\approx1.089$.
\end{itemize}
The second case is a transcendental analogue of the single-horizon case $f_0\le0<f_\infty$ of Tab.~\ref{tab:bell-broad}: a single horizon beyond the core from a genuinely non-monotonic, non-rational $f$ ; note that a Gaussian $f_\gamma$ cannot realize the exotic pattern $f_0>0>f_\infty$, as shown above.

\subsection{Algebraic type and energy conditions}
\label{sec:model-check}

\begin{proposition}[Algebraic type across all models]
\label{prop:model-algebraic}
For any of the families Eqs.~\eqref{eq:model-sigmoid}, \eqref{eq:model-bell}, \eqref{eq:model-cubic}, \eqref{eq:model-hayward}, \eqref{eq:model-tanh}, \eqref{eq:model-logistic}, \eqref{eq:model-gauss} — rational or not — paired with an affine $\Sigma$ — either a one-sided boundary $\Sigma=\Sigma_0+r$ with $\Sigma_0>0$, or the point-like-center gauge $\Sigma(r)=r$ (the case $\Sigma_0=0$) — the source is of the isotropic Segre type $[(1,1)(11)]$: $\rho+p_r\equiv0$, in agreement with Thm.~\ref{thm:affine}. Paired instead with a throat, $\Sigma=\sqrt{r^2+\Sigma_0^2}$, the source is of the generic type $[1,1(11)]$, with
\begin{equation}
\label{eq:model-nec-radial}
    (\rho+p_r)\big|_{r=0} = -\frac{f(0)}{4\pp\Sigma_0^2} .
\end{equation}
\end{proposition}

\begin{proof}
Both $\Sigma=\Sigma_0+r$ and $\Sigma(r)=r$ are the affine form Eq.~\eqref{eq:affine-sigma} (with $\Sigma_0>0$ and $\Sigma_0=0$ respectively), so $\Sigma''\equiv0$ and Eq.~\eqref{eq:nec-radial} gives $\rho+p_r\equiv0$ for any analytic $f$ — rational or transcendental, the computation never differentiates $f$ itself — matching Thm.~\ref{thm:affine}; this is the gauge-theoretic reason the classical RBH literature, built almost exclusively in the gauge $\Sigma=r$, treats $p_r=-\rho$ as a standing assumption rather than a special case. For the throat $\Sigma=\sqrt{r^2+\Sigma_0^2}$, $\Sigma''(0)=1/\Sigma_0\neq0$, so Eq.~\eqref{eq:nec-radial} gives $(\rho+p_r)|_{r=0}=-f(0)\Sigma''(0)/(4\pp\Sigma_0)=-f(0)/(4\pp\Sigma_0^2)$, vanishing exactly when $f(0)=0$ (Prop.~\ref{prop:nec-marginal}) and generically not otherwise, confirming the source is not isotropic there (Cor.~\ref{cor:nothroat}).
\end{proof}

\begin{proposition}[NEC/SEC at the throat models]
\label{prop:model-nec-sec}
For the throat pairing of Sec.~\ref{sec:model-sigmoid-beyond}, $(\rho+p_r)(0)=1/(4\pp a^2)>0$ (since $f_\sigma(0)=-1$ there). For the throat pairing of Sec.~\ref{sec:model-sigmoid-core} and Sec.~\ref{sec:model-bell-core}, $(\rho+p_r)(0)=0$ exactly, since $f(0)=0$ in each (Prop.~\ref{prop:nec-marginal}).
\end{proposition}

\begin{proof}
Direct substitution into Eq.~\eqref{eq:model-nec-radial}.
\end{proof}

\subsection{Comparison of model parameters}
\label{sec:model-compare}

Table~\ref{tab:models-final} collects every explicit model of this section.

\begin{table}[ht]
\centering
\setlength{\tabcolsep}{3.5pt}
\begin{tabular}{llccccc}
\hline
Model & Shape & Rational? & $r=0$ type & Horizons & $S/\pp$ & Segre type \\
\hline
\S\ref{sec:model-sigmoid-beyond}, throat & sigmoid & yes & throat & $1$ & $1+a^2$ & $[1,1(11)]$ \\
\S\ref{sec:model-sigmoid-beyond}, one-sided & sigmoid & yes & one-sided & $1$ & $1+a^2$ & $[1,1(11)]$ \\
\S\ref{sec:model-sigmoid-core}, one-sided & sigmoid & yes & one-sided & $1$ (core) & $\Sigma_0^2$ & $[(1,1)(11)]$ \\
\S\ref{sec:model-sigmoid-core}, throat & sigmoid & yes & throat & $1$ (core) & $\Sigma_0^2$ & $[1,1(11)]$ \\
\S\ref{sec:model-sigmoid-point} & sigmoid & yes & point-like & $1$ (exotic) & $r_h^2$ & $[(1,1)(11)]$ \\
\S\ref{sec:model-sigmoid-transcendental}, $f_\tau$ & sigmoid & no & one-sided & $1$ & $(\Sigma_0+r_h)^2$ & $[(1,1)(11)]$ \\
\S\ref{sec:model-sigmoid-transcendental}, $f_\ell$ & sigmoid & no & throat/one-sided & $1$ (core) & $\Sigma_0^2$ & mixed$^*$ \\
\S\ref{sec:model-bell-core}(a) & bell & yes & throat/one-sided & $2$ & $\Sigma(r_+)^2$ & mixed$^*$ \\
\S\ref{sec:model-bell-core}(b) & bell & yes & throat/one-sided & $1$ (exotic) & $\Sigma(r_h)^2$ & mixed$^*$ \\
\S\ref{sec:model-hayward} & bell & yes & point-like & $2$ (classical) & $r_+^2$ & $[(1,1)(11)]$ \\
\S\ref{sec:model-bell-transcendental}, $N=2$ & bell & no & throat/one-sided & $2$ & $\Sigma(r_+)^2$ & mixed$^*$ \\
\S\ref{sec:model-bell-transcendental}, $N=1$ & bell & no & throat/one-sided & $1$ & $\Sigma(r_h)^2$ & mixed$^*$ \\
\hline
\end{tabular}
\captionsetup{width=.9\textwidth}
\caption{All explicit models of Sec.~\ref{sec:model}. $^*$Isotropic for the one-sided (affine) pairing, generic type $[1,1(11)]$ for the throat pairing (Prop.~\ref{prop:model-algebraic}). Horizon counts refer to the patch $r\ge0$; the doubling of the horizon under the mirror extension $r\to|r|$ requires an even analytic $f$ and does not occur for the sigmoid models above (Sec.~\ref{sec:model-sigmoid-beyond}). $S/\pi=\Sigma(r_h)^2$ at the (outer) horizon, Eq.~\eqref{eq:temp-entropy}, with $\Sigma(r_h)=r_h$ for $\Sigma=r$, $\Sigma_0+r_h$ for the one-sided and $\sqrt{r_h^2+\Sigma_0^2}$ for the throat pairing.}
\label{tab:models-final}
\end{table}

Four patterns are worth drawing out. First, within each single-horizon mechanism, the choice of throat versus one-sided boundary changes the algebraic type (isotropic only for the affine, one-sided choice, by Thm.~\ref{thm:affine}/Cor.~\ref{cor:nothroat}) but not the horizon's local invariants ($\kappa$, $S$) — a purely kinematic fact already established in Prop.~\ref{prop:model-Ap}. Second, the horizon-at-the-core mechanism always yields the mass-independent entropy floor $S=\pp\Sigma_0^2$ of Rem.~\ref{rem:min-entropy}, in contrast to the mass-dependent $S$ of the beyond-the-core and classical double-horizon mechanisms. Third, exactly one non-degenerate horizon at a point-like center is realizable, but only by abandoning the asymptotically static hypothesis ($f_\infty<0$, Rem.~\ref{rem:point-center-single}); every other single-horizon row instead uses a finite-area core, where no such trade-off is required. Fourth, rationality of $f$ plays no role anywhere in the table: the transcendental rows (6, 7, 11, 12) reproduce every qualitative feature of their rational counterparts, exactly as Prop.~\ref{prop:sigmoid-count} and Thm.~\ref{thm:bell-count} predict, since neither result's proof uses more than analyticity and monotonicity.

The analysis so far is confined to the patch $r\in[0,\infty)$,
on which the single-horizon models have no inner Cauchy horizon.
Removing the inner horizon raises a further question:
where does an infalling observer end up?
We end this section by discussing this issue.
Consider a finite-area core.
A radially infalling observer with $E^{2}>f(0)$ (every observer if $f(0)\le0$)
reaches $r=0$ in finite proper time.
If $r=0$ is a horizon (Thm.~\ref{thm:horizonatthroat}), the observer crosses a
regular Killing horizon.
Otherwise $r=0$ is a hypersurface of finite curvature at which the patch ends:
it is spacelike inside the $T$-region if $f(0)<0$
(see Rem.~\ref{rem:beyondcore}) and timelike inside a static region if $f(0)>0$.
The patch is therefore geodesically incomplete, and the observer's fate depends
on how the geometry continues to $r<0$.

\begin{figure}[!htbp]
	\centering
	\begin{subfigure}[b]{0.333\textwidth}
		\centering
		\includegraphics[width=\textwidth]{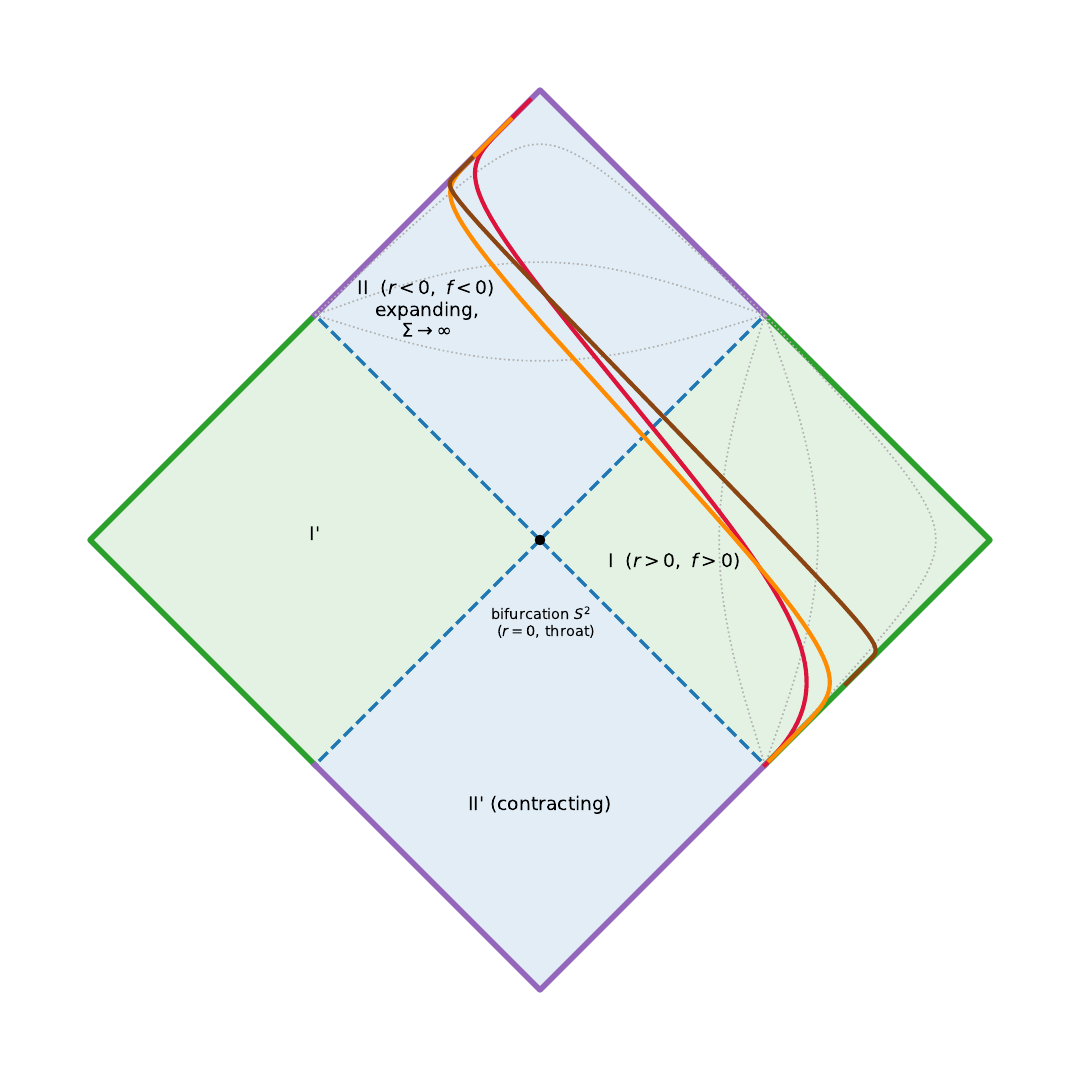}
		\caption{}
	\end{subfigure}
	\hspace{-3mm}
	\begin{subfigure}[b]{0.333\textwidth}
		\centering
		\includegraphics[width=\textwidth]{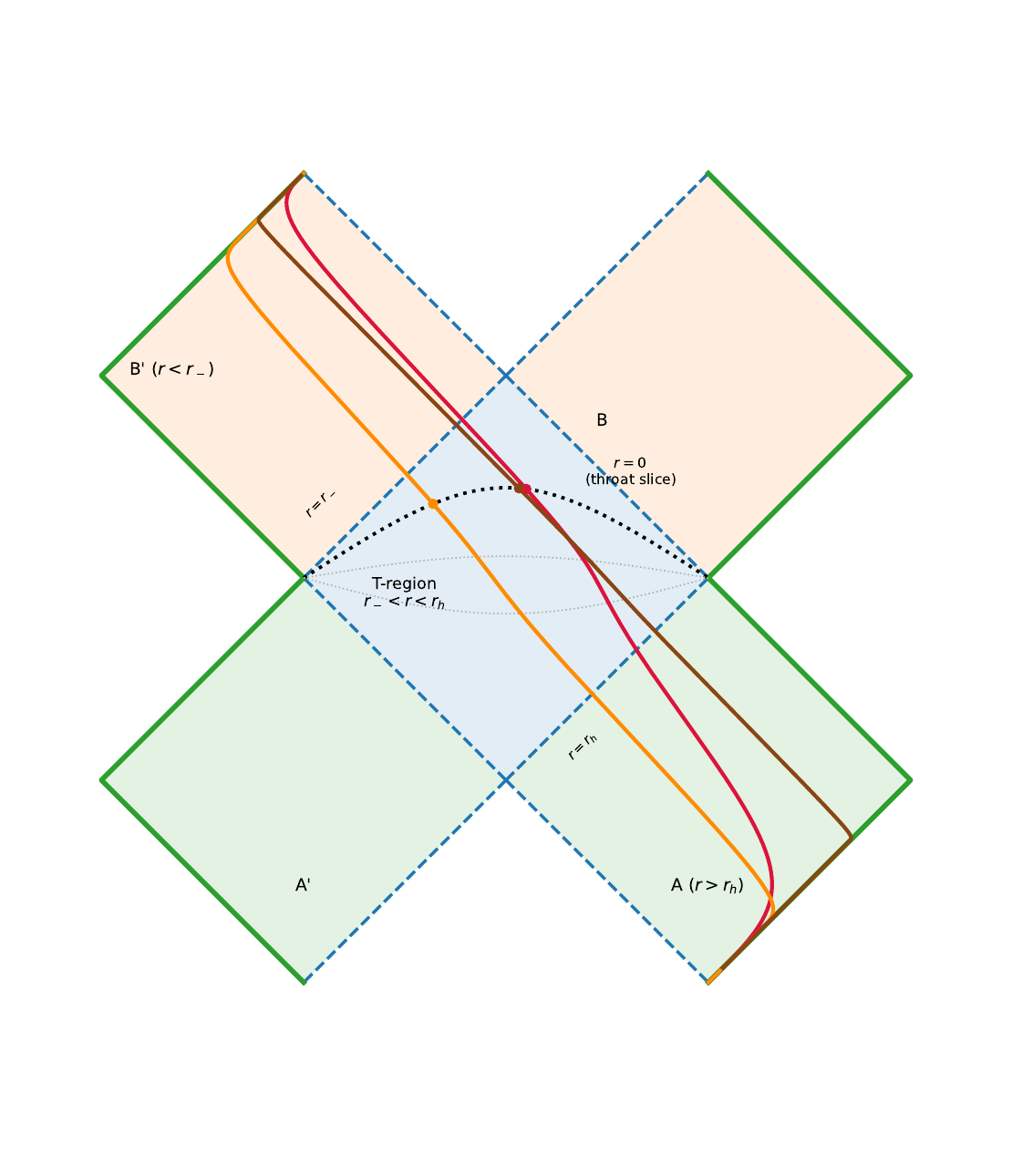}
		\caption{}
	\end{subfigure}
	\hspace{-3mm}
	\begin{subfigure}[b]{0.333\textwidth}
		\centering
		\includegraphics[width=\textwidth]{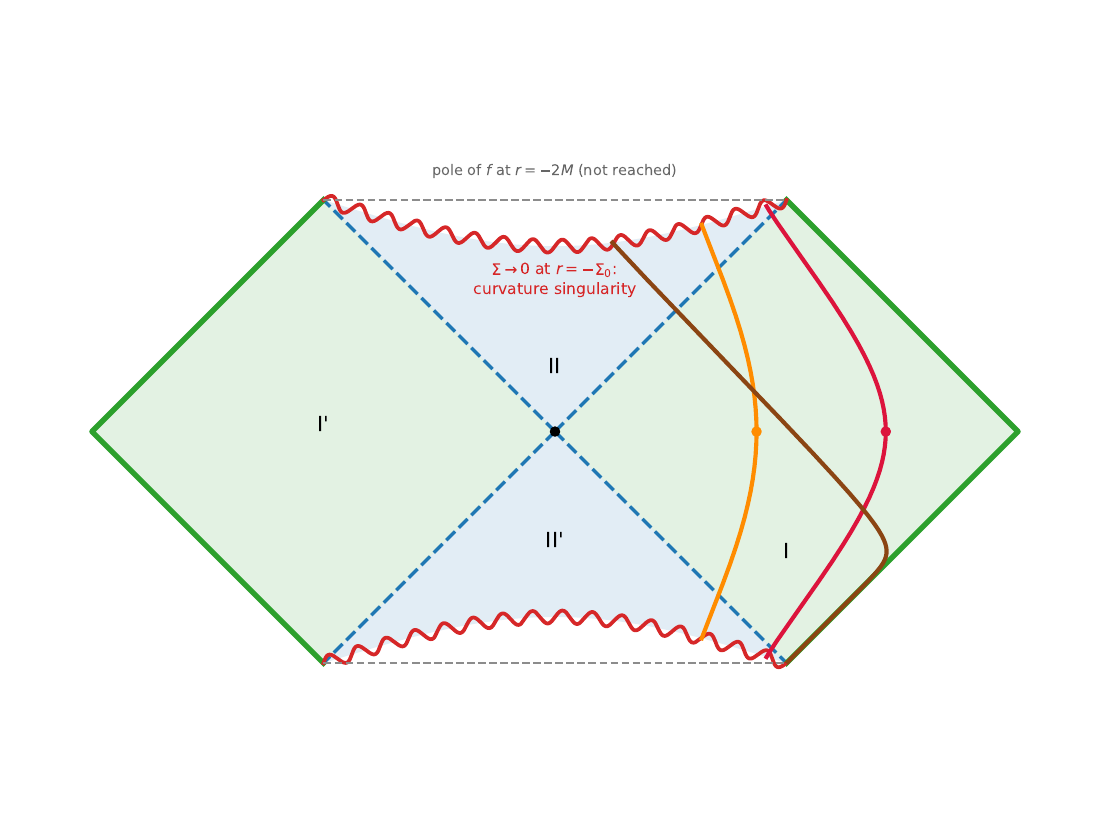}
		\caption{}
	\end{subfigure}
	\begin{subfigure}[b]{0.333\textwidth}
		\centering
		\includegraphics[width=\textwidth]{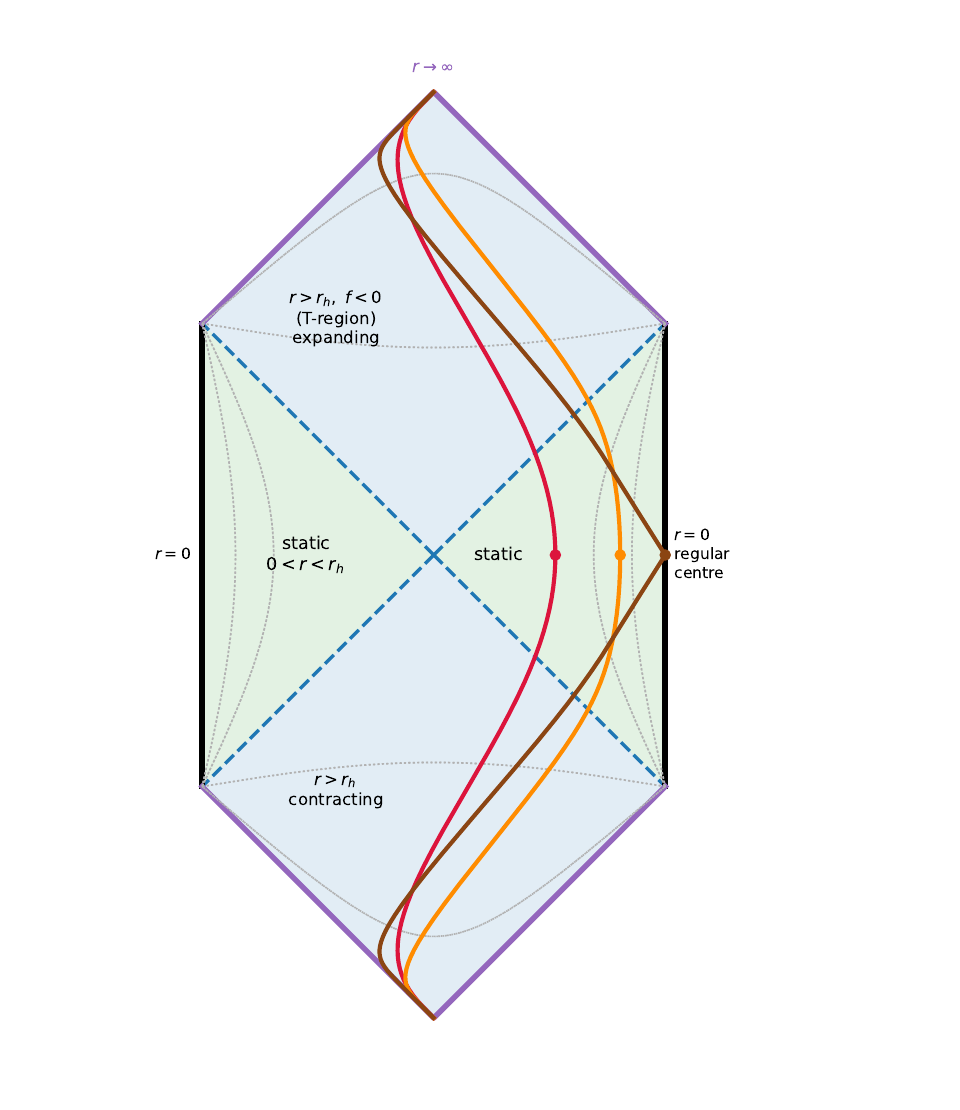}
		\caption{}
	\end{subfigure}
	\hspace{-3mm}
\begin{subfigure}[b]{0.333\textwidth}
		\centering
		\includegraphics[width=\textwidth]{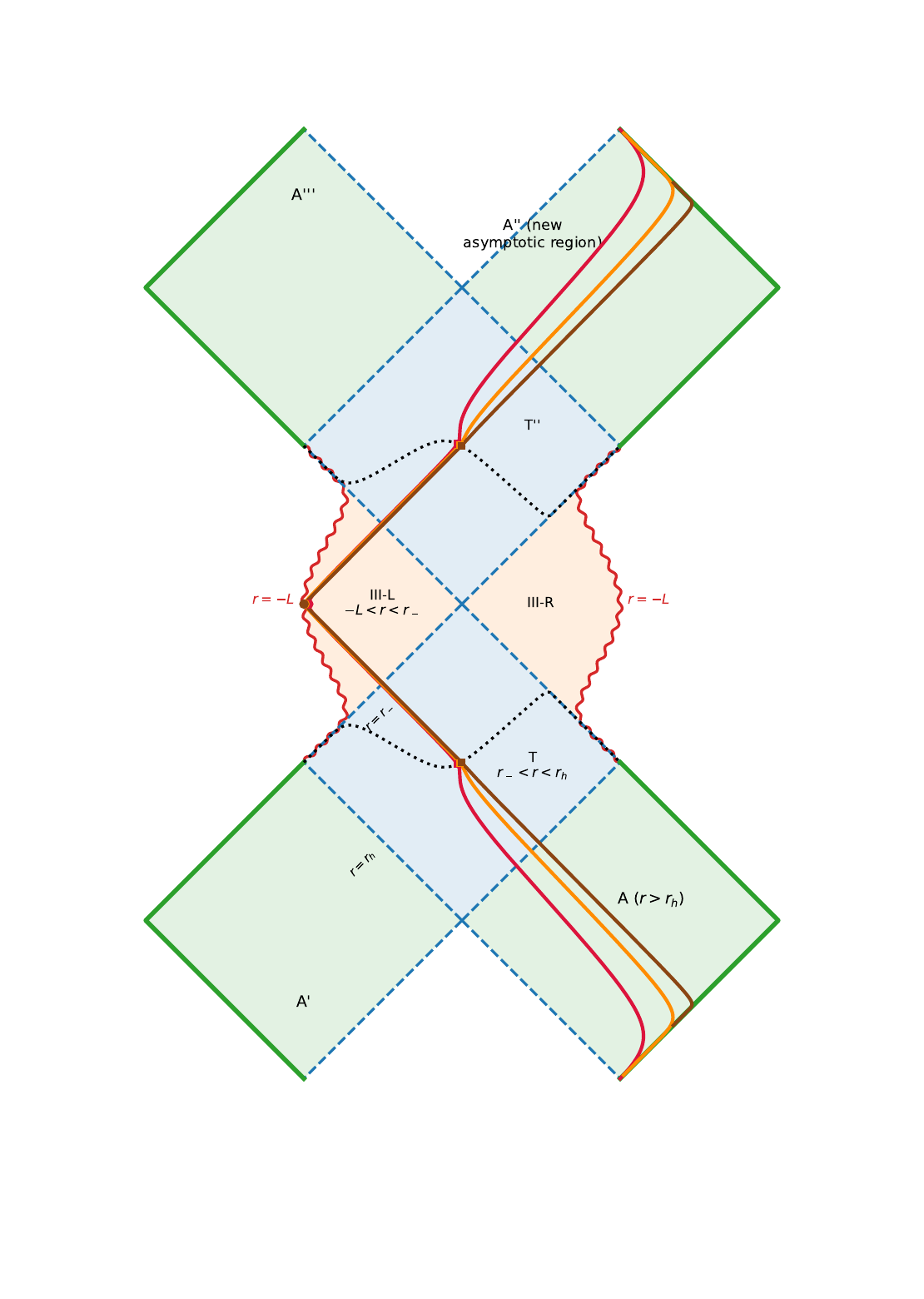}
		\caption{}
	\end{subfigure}
	\hspace{-3mm}
	\begin{subfigure}[b]{0.333\textwidth}
		\centering
		\includegraphics[width=\textwidth]{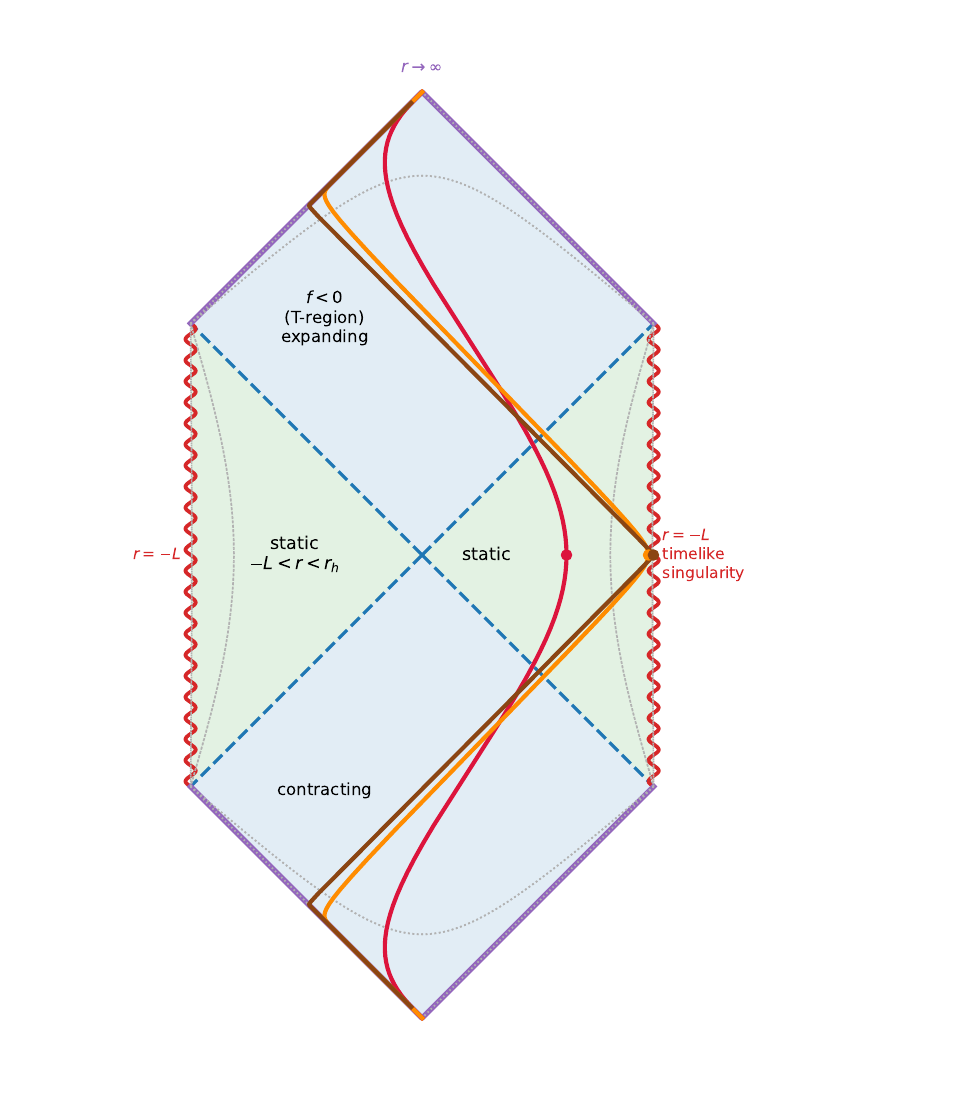}
		\caption{}
	\end{subfigure}
	\captionsetup{width=.9\textwidth}
  \caption{Two-dimensional causal diagrams of analytic continuations of
  single-horizon patches. 
  (a)~Endless expansion: horizon at a throat, $f=\tanh(r/2M)$, $\Sigma=\sqrt{r^{2}+\Sigma_0^{2}}$;
  the observer moves toward $r\to-\infty$ (purple), reached only after infinite proper time.
  (b)~Second asymptotic region: $f=1-4\me^{-4(r-1/2)^{2}}$ with a throat; the observer crosses
  $r_h$, 
  the regular slice $r=0$ (dotted) and the horizon $r_-$. 
  (c)~Spacelike singularity: pole of $f$ at $r=-L$, or $\Sigma\to0$ for a one-sided $\Sigma$ (red).
  (d)~Point-like center, $f=(1-r^{3})/(1+r^{3})$, $\Sigma=r$: the horizon at $r_h=1$ is cosmological and the static regions $0<r<r_h$ contain the regular centers (black).
  (e)~RN-like continuation: 
  after $r_h$ comes an
  inner horizon $r_-<0$ and then a timelike singularity at $r=-L$.
  (f)~Cosmological horizon followed by a timelike singularity at $r=-L$.
  Dashed blue lines are Killing horizons, green lines are null infinity, the dotted
  line is the slice $r=0$, and the colored curves are radial timelike geodesics with different energies $E$ (red wavy lines: curvature singularities; squares: crossings of $r=0$; dots: release or bounce points).
  }
  \label{fig:penrose}
\end{figure}

We continue $f$ and $\Sigma$ analytically to $r<0$ along the real axis.
This continuation is natural but not unique:
the matter content at $r<0$ is not fixed by the analysis above.
Near each simple zero of $f$ we introduce Kruskal coordinates.
Where $f\neq0$, the $(t,r)$ part of the metric Eq.~\eqref{eq:metric} is
conformal to two-dimensional Minkowski space with $dr_*=dr/f$, which suffices to
draw the diagrams in Fig.~\ref{fig:penrose}.
These are two-dimensional causal diagrams, not conformal compactifications of
the four-dimensional spacetime.
A radial timelike geodesic with energy $E$ obeys $\dot r^{2}=E^{2}-f(r)$,
and we use this equation to decide whether an endpoint is reached in finite
proper time.

The continuation leads to five classes of fate, which do not coincide with the two classes
of Sec.~\ref{sec:theory-single}.
The class is decided by the behavior of $f$ at $r<0$ (a pole, a further zero, or
neither) and by the form of $\Sigma$ (a throat or one-sided).

\emph{Endless expansion.}
Suppose $f$ is real analytic on $\mathbb R$ with $f$, $f'$, $f''$ bounded and
simple zeros only, and $\Sigma=\sqrt{r^{2}+\Sigma_0^{2}}$.
Then $A$, $B$, $C$, $D$ in Eq.~\eqref{eq:ABCDdef} are bounded, because
$\Sigma\ge\Sigma_0>0$, $\Sigma'/\Sigma=r/(r^{2}+\Sigma_0^{2})$ and
$\Sigma''=\Sigma_0^{2}/\Sigma^{3}$ are bounded.
Moreover $\dot r^{2}\le E^{2}+\sup|f|$, so a radial observer needs infinite
proper time to reach $r\to\pm\infty$; and since $|f|$ is bounded, $\int dr/|f|$
diverges, so these ends are the null edges (with their corner) drawn in purple,
not spacelike lines.
Suppose also $f<0$ for all $r<r_h$, as for $f_\ell=\tanh(r/2M)$ of
Sec.~\ref{sec:model-sigmoid-transcendental} and for $f_\tau$ of
Eq.~\eqref{eq:model-tanh} with $f_0<0<f_\infty$.
The causal structure is then that of Fig.~\ref{fig:penrose}(a).
The observer crosses the future horizon, passes the sphere of minimal area, and
moves toward $r\to-\infty$ in a region where $f<0$ and $\Sigma$ grows without
bound.
No singularity is met and no inner horizon bounds this region.

\emph{Passage to a second asymptotic region.}
If $f$ is as above except that it has a second simple zero $r_-<0$ and tends to a
positive constant as $r\to-\infty$, two horizons bound the $T$-region.
The Gaussian-dip model of Sec.~\ref{sec:model-bell-transcendental} with $f_\infty=1$,
$f_{\min}=-3$, $r_*=w=1/2$ and a throat is an example: it has simple zeros at
$r_h\simeq1.089$ and $r_-\simeq-0.089$.
The observer crosses $r_h$, passes the throat, and leaves through the horizon at
$r_-$ into a second asymptotically flat region Fig.~\ref{fig:penrose}(b). For $E>0$ one has $\dot t=E/f<0$ in the $T$-region, so the observer drifts to the left and leaves through the left branch of $r_-$ (exterior drawn on the right), not through the right branch ($v\to\infty$) at which mass inflation is expected.
This structure resembles a black bounce, but it requires no mirror symmetry and
no matter shell.
The horizon at $r_-$ has $\kappa_-=|f'(r_-)|/2>0$ and bounds the $T$-region to the
future, as the inner horizon does in a two-horizon RBH.
It lies outside the patch, but the argument of Prop.~\ref{prop:massinflation}
suggests that mass inflation may reappear there.

\emph{Spacelike singularity.}
For the rational family $f_\sigma$ of Eq.~\eqref{eq:model-sigmoid}, or for a
one-sided $\Sigma=\Sigma_0+r$, the continuation reaches a curvature
singularity at finite proper time.
The function $f_\sigma$ has a simple pole at $r=-L$, where $A\to\infty$.
For $\Sigma=\Sigma_0+r$, $\Sigma$ vanishes at $r=-\Sigma_0$, where
$D=(1-f\Sigma'^{2})/\Sigma^{2}$ diverges unless $f-1=O(\Sigma^{2})$, which fails
here since $f<0$.
In these cases $f<0$ at the singularity, so it is spacelike, and the diagram is
Schwarzschild-like Fig.~\ref{fig:penrose}(c).
Thus removing the inner horizon removes the mass-inflation mechanism, but it does
not by itself remove the singularity: in this continuation the singularity is
displaced behind the horizon.

\emph{de Sitter-like.}
For the point-like-center model of Sec.~\ref{sec:model-sigmoid-point}, the static region
is $0<r<r_h$ and contains the regular center, while $r>r_h$ is a time-dependent
region with $f\to-1$.
The horizon is of cosmological type, and the diagram is a hexagon with a regular
center on each side Fig.~\ref{fig:penrose}(d).
An observer who leaves the static region reaches $r\to\infty$ only after infinite
proper time.
In the static region $f'<0$, so Thm.~\ref{thm:V-positive} does not apply there.

\emph{Timelike singularity.}
The family $f_\beta$ of Eq.~\eqref{eq:model-bell} is the Reissner--Nordstr\"om
lapse in $\rho=r+L$, so its continuation ends on a timelike singularity at $r=-L$,
where $f_\beta\to+\infty$.
If $f_0<0<f_\infty$,
the horizon $r_h$ is followed by a second horizon
$r_-=\rho_--L<0$, with $\rho_-$ the smaller root in Eq.~\eqref{eq:bell-horizons},
 before the singularity is reached. The structure is RN-like Fig.~\ref{fig:penrose}(e): $r_-$ is a Cauchy horizon with $\kappa_->0$, to which the argument of Prop.~\ref{prop:massinflation} applies. A radial observer with $E>1$ coming from the exterior enters the left region beyond $r_-$, is repelled by the singularity ($f\to+\infty$) and turns around at $f(r_t)=E^2$ without reaching it, and re-emerges through a second $T$-region into a new asymptotic region Fig.~\ref{fig:penrose}(e).
If $f_0>0>f_\infty$, 
there is no second
horizon: the static region $-L<r<r_h$ contains the core, $r=0$ is not a regular
center, and the region ends on the timelike singularity Fig.~\ref{fig:penrose}(f). Radial timelike geodesics turn around at $f(r_t)=E^2$ and never reach $r=-L$; radial null geodesics do reach it at finite affine parameter.

A mirror-symmetric extension $r\to|r|$ is also possible when $r=0$ is a throat, but it is
real analytic only if $f'(0)=0$ (proof of Cor.~\ref{cor:noextension}).
If $f'(0)\neq0$, as for all the sigmoid models of Sec.~\ref{sec:model} paired with a throat, it has a kink at $r=0$ and
produces a $\delta$-function layer, since $G^{\theta}{}_{\theta}\propto f''$:
a shell with vanishing surface density and $\delta$-function tangential pressure.
We do not regard such a spacetime as regular, and it is not drawn in
Fig.~\ref{fig:penrose}.
This applies to $f_\sigma$ of Sec.~\ref{sec:model-sigmoid-beyond}: the doubling of its
horizon, which Prop.~\ref{prop:mirror} establishes only for even analytic $f$ (as in the
black-bounce lapse $1-2m/\sqrt{r^{2}+a^{2}}$), occurs only in this non-regular extension.

Thus the absence of an inner horizon on the patch does not fix the global fate of
the observer.
In the endless-expansion, spacelike-singular and de Sitter-like continuations, and in the
cosmological continuation ending on a timelike singularity Fig.~\ref{fig:penrose}(f),
the extended region beyond $r_h$ contains no further Killing horizon with $\kappa>0$, and the
blueshift mechanism of Prop.~\ref{prop:massinflation} has nothing to act on there.
In the second-asymptotic-region and RN-like continuations, the horizon at $r_-$ may restore it.
Since the continuation is not unique, the statements on mass inflation in
Sec.~\ref{sec:stability} refer to the patch $r\ge0$.

\section{Stability}
\label{sec:stability}

Sections~\ref{sec:theory}--\ref{sec:model} determined which horizon structures the metric in Eq.\eqref{eq:metric} admits and provided explicit realizations of each. We now ask how these spacetimes respond to linear perturbations, and in particular whether removing the inner (Cauchy) horizon, the single-horizon mechanisms of Sec.\ref{sec:theory-single}, also removes the classical instability associated with it. We show this in two separate steps: first, that the specific mechanism responsible for the instability of ordinary (point-like-center, two-horizon) RBHs is structurally absent when there is only one horizon; second, and independently, that the resulting spacetimes are also free of the standard linear mode instabilities that the mere absence of mass inflation does not by itself rule out.

\subsection{Master equation and effective potential}
\label{sec:stability-setup}

We probe the dynamics of metric Eq.~\eqref{eq:metric} with a test scalar field $\Phi$, $\Box\Phi=0$, the standard proxy for the qualitative stability properties of such spacetimes. Separating
\begin{equation}
    \Phi = \frac{\Psi(t,r)}{\Sigma(r)}\,Y_{lm}(\theta,\phi) ,
\end{equation}
the wave equation $\frac{1}{\sqrt{-g}}\partial_\mu(\sqrt{-g}\,g^{\mu\nu}\partial_\nu\Phi)=0$, with $\sqrt{-g}=\Sigma^2\sin\theta$, $g^{tt}=-1/f$, $g^{rr}=f$, $g^{\theta\theta}=1/\Sigma^2$, reduces after separating the angular part (eigenvalue $-l(l+1)$) to the radial equation
\begin{equation}
    \omega^2 R + \frac{f}{\Sigma^2}\frac{\dif}{\dif r}\!\left(\Sigma^2 f\,\frac{\dif R}{\dif r}\right) - \frac{f\,l(l+1)}{\Sigma^2}R = 0 , \qquad \Phi\propto R(r)e^{-i\omega t}Y_{lm} .
\end{equation}

\begin{proposition}[Master equation]
\label{prop:master}
Let $r_*$ be the tortoise coordinate, $\dif r_*/\dif r = 1/f$, and set $\Psi:=\Sigma R$. Then
\begin{equation}
\label{eq:master}
    \frac{\dif^2\Psi}{\dif r_*^2} + \big[\omega^2-V(r)\big]\Psi = 0 , \qquad
    V(r) = f(r)\left[\frac{l(l+1)}{\Sigma(r)^2}+\frac{1}{\Sigma(r)}\frac{\dif}{\dif r}\Big(f(r)\Sigma'(r)\Big)\right] .
\end{equation}
\end{proposition}

\begin{proof}
Writing $R=\Psi/\Sigma$ in the radial equation above and multiplying through by $\Sigma$ gives, after collecting terms (a direct if lengthy computation, most transparently organized by first expanding $\frac{f}{\Sigma^2}\frac{\dif}{\dif r}(\Sigma^2 f R')$ and then substituting $R=\Psi/\Sigma$),
\begin{equation}
    f\frac{\dif}{\dif r}\!\left(f\frac{\dif\Psi}{\dif r}\right) + \omega^2\Psi - f\left[\frac{l(l+1)}{\Sigma^2}+\frac{(f\Sigma')'}{\Sigma}\right]\Psi = 0 .
\end{equation}
Since $\dif/\dif r_* = f\,\dif/\dif r$ by definition of $r_*$, the first term is exactly $\dif^2\Psi/\dif r_*^2$, giving Eq.~\eqref{eq:master} with $V$ as stated.
\end{proof}

\begin{remark}
For $f=1-2M/r,\ \Sigma=r$ (Schwarzschild), Eq.~\eqref{eq:master} gives $V=(1-2M/r)\big[l(l+1)/r^2+2M/r^3\big]$, the standard Regge--Wheeler potential for a spin-0 field, as it must.
\end{remark}

On the static exterior region where $f>0$, we impose the standard quasinormal boundary conditions: purely outgoing as $r_*\to+\infty$ (or towards whatever asymptotic region is relevant) and purely ingoing at the horizon ($r_*\to-\infty$). A quasinormal frequency $\omega=\omega_R+i\omega_I$ is unstable if $\omega_I>0$ (exponential growth in time, since $\Psi\propto \me^{-i\omega t}=\me^{-i\omega_Rt+\omega_It}$) and stable if $\omega_I\le0$ for every $l$.

\subsection{Mass inflation of double-horizon RBHs}
\label{sec:stability-massinflation}

We first treat the center-regular class of Sec.~\ref{sec:theory-center} in the physically standard, asymptotically static case $f_\infty>0$: by Prop.~\ref{prop:exact2}(b), a genuine (unimodal, $f_{\min}<0$) such BH has two horizons, $0<r_-<r_+$, non-degenerate when $f'\neq0$ there (as for all explicit models of Sec.~\ref{sec:model}), with surface gravities $\kappa_\pm=|f'(r_\pm)|/2$ (Def.~\ref{def:horizon}); $\kappa_->0$ since $r_-$ is non-degenerate. The region $r_-<r<r_+$ is, as in Rem.~\ref{rem:beyondcore}, a non-static $T$-region, and $r=r_-$ is a Cauchy horizon: the boundary of the domain of dependence of data given on a spacelike slice of the exterior.

\begin{proposition}[Mass inflation at $r_-$]
\label{prop:massinflation}
Let $0<r_-<r_+$ be as above, with $\kappa_->0$. Any generic perturbation incident on the exterior produces, upon propagating into the region $r_-<r<r_+$ and approaching $r_-$, a local mass function (equivalently, the Kretschmann scalar $K$) that grows without bound, at a rate set by $\kappa_-$:
\begin{equation}
\label{eq:mass-inflation}
    m(v) \sim m_0\,\exp(2\kappa_- v) \times \big(\text{power-law tail}\big) \longrightarrow \infty \qquad (v\to\infty) .
\end{equation}
Here the power-law tail is the late-time behaviour $T_{vv}\sim v^{-p}$ ($p>0$, $v$ advanced time) that generically remains after the initial transient of any perturbation has decayed — a general feature of wave propagation on any asymptotically flat background whose effective potential falls off as $l(l+1)/r^2$ at large $r$. Consequently, every center-regular RBH satisfying the hypotheses of Prop.~\ref{prop:exact2}(b) with $f_\infty>0$ is unstable at its inner horizon.
\end{proposition}

\begin{proof}[Proof sketch]
Near $r=r_-$, ingoing Eddington--Finkelstein coordinates $(v,r)$, $v=t+r_*$, give $\dif s^2=-f\,\dif v^2+2\,\dif v\,\dif r+\Sigma^2\dif\Omega^2$, regular across the branch of $r_-$ reached at finite $v$ (the left branch, with the exterior drawn on the right); the Cauchy-horizon branch, on which the blueshift acts, is instead reached as $v\to\infty$ at finite retarded time $u=t-r_*$, where the outgoing coordinates $(u,r)$ are regular, and there $r-r_-\propto\me^{-\kappa_-(v-u)}$ because $r_*=(v-u)/2\to+\infty$ at $r_-$. This is precisely the geometric origin of the blueshift: an observer crossing the Cauchy-horizon branch at finite $u$ receives the entire infinite advanced-time history $v\to\infty$ of the ingoing flux, and in the horizon-regular coordinate $V=-\me^{-\kappa_- v}$ a flux component $T_{vv}$ is amplified by the factor $(\dif v/\dif V)^2\propto\me^{2\kappa_- v}$. Since the Misner--Sharp mass obeys $\partial_v m\propto \Sigma^2 f\,T_{vv}$ along ingoing rays (its precise coefficient and the accompanying outgoing-flux equation are given in App.~\ref{app:self_energy}), the mass accumulated by advanced time $v$ from a source with the power-law tail above satisfies an integral of the schematic form $\delta m(v)\propto\int^v \me^{2\kappa_- v'}(v')^{-p}\dif v'$, which diverges as $v\to\infty$ for any $p>0$: exponential blueshift always defeats power-law decay. This is the Poisson--Israel mass-inflation mechanism; Eq.~\eqref{eq:mass-inflation} is its parametric signature, with full quantitative derivation deferred to App.~\ref{app:self_energy}. Since $K\sim m^2/\Sigma^6$ near $r_-$ by the Kretschmann formula Eq.~\eqref{eq:K} (specializing to the point-like-center gauge, where $D=2m/\Sigma^3$ dominates near a horizon), the divergence of $m$ signals divergent curvature as measured by an infalling observer approaching $r_-$ — a genuine, gauge-invariant instability, not a coordinate artifact.
\end{proof}

\begin{remark}
The mechanism requires only $\kappa_->0$ (a non-degenerate inner horizon) and a late-time tail at the outer horizon that is not exponentially suppressed; both are generic. The former follows from the non-degenerate two-horizon assumption in Prop.\ref{prop:exact2}(b), while the latter follows from Price’s theorem \cite{Price:1972pw,Gundlach:1993tp,Dafermos:2005eh}, which depends only on the common large-$r$ falloff $V\sim l(l+1)/r^2$ implied by Assump.\ref{as:flat}. It therefore applies to every model in this class, rather than being a special feature of any particular model (such as the Hayward model of Sec.~\ref{sec:model-hayward}).
\end{remark}

\subsection{Absence of the mass-inflation in single-horizon RBHs}
\label{sec:stability-absence}

Mass inflation, as derived above, is not a generic feature of perturbed BHs: it requires specifically a second, inner Killing horizon behind the outer one, across which the blueshift mechanism of Prop.~\ref{prop:massinflation} operates. We show that neither single-horizon mechanism of Sec.~\ref{sec:theory-single} provides one.

\begin{proposition}[No Cauchy horizon at a core-coincident horizon]
\label{prop:no-cauchy-atcore}
In the setting of Thm.~\ref{thm:horizonatthroat}, the patch $[0,\infty)$ contains no point at which the blueshift mechanism of Prop.~\ref{prop:massinflation} can operate.
\end{proposition}

\begin{proof}
By Thm.~\ref{thm:horizonatthroat}(iii), $f>0$ throughout $r>0$, and $r=0$ is the only zero of $f$ on $[0,\infty)$. Thus, the patch contains no sub-horizon region and hence no second horizon or Cauchy horizon. The mass-inflation mechanism requires two horizons bounding a non-static region between them and therefore cannot operate here.
\end{proof}

\begin{proposition}[No Cauchy horizon beyond the core]
\label{prop:no-cauchy-beyond}
In the setting of Lem.~\ref{lem:singlebeyond}, the region $0\le r<r_h$ is bounded, on the patch, by the single horizon $r_h$ on one side and the (regular!) core $r=0$ on the other, not by a second horizon.
\end{proposition}

\begin{proof}
By Lem.~\ref{lem:singlebeyond}, $f$ has exactly one zero on $[0,\infty)$, at $r_h$; hence $f\neq0$ throughout $[0,r_h)$, and in particular $r=0$ is not a horizon (it is a finite-area core, Prop.~\ref{prop:throatreg} having already established its regularity independently of the sign of $f(0)$). The non-static region $0\le r<r_h$ (Rem.~\ref{rem:beyondcore}) therefore ends, at its inner end, at the finite-curvature hypersurface $r=0$ of the patch rather than at a second Killing horizon. Whether a second horizon appears after continuing through $r=0$ depends on the continuation, which is not unique (Sec.~\ref{sec:model-compare}).
\end{proof}

Together, Props~\ref{prop:no-cauchy-atcore} and~\ref{prop:no-cauchy-beyond} give the first, structural half of the stability result:

\begin{quote}
Level 1. On its patch $r\ge0$, every single-horizon RBH of Sec.~\ref{sec:theory-single} — whether the horizon sits at or beyond the core — lacks the Cauchy horizon required for the Poisson--Israel mass-inflation instability of Prop.~\ref{prop:massinflation}: the mechanism responsible for the classical instability of ordinary RBHs is structurally absent on the patch, not merely quantitatively suppressed. Whether an inner horizon can reappear beyond the patch depends on the continuation (Sec.~\ref{sec:model-compare}).
\end{quote}

This is weaker than a proof of stability: the absence of one particular instability channel does not by itself preclude others, a point we take up next.

\subsection{Dynamical mode stability}
\label{sec:stability-modestability}

Level 2 requires testing the single-horizon spacetimes for ordinary linear mode instabilities directly, via the master equation Eq.~\eqref{eq:master}. We use the standard energy method.

\begin{proposition}[Positive potential precludes growing modes]
\label{prop:energy-method}
If $V(r_*)\ge0$ throughout the domain of Eq.~\eqref{eq:master}, no solution with $\omega_I>0$ exists compatible with finite-energy (compactly supported, or suitably decaying) initial data.
\end{proposition}

\begin{proof}
For real $\Psi(t,r_*)$ solving $\partial_t^2\Psi=\partial_{r_*}^2\Psi-V\Psi$, define $E(t)=\tfrac12\int\big[(\partial_t\Psi)^2+(\partial_{r_*}\Psi)^2+V\Psi^2\big]\dif r_*$. Then
\begin{equation}
    \frac{\dif E}{\dif t} = \int\Big[\partial_t\Psi\,\partial_t^2\Psi+\partial_{r_*}\Psi\,\partial_t\partial_{r_*}\Psi+V\Psi\,\partial_t\Psi\Big]\dif r_* = \int\partial_{r_*}\!\big(\partial_t\Psi\,\partial_{r_*}\Psi\big)\dif r_* = \Big[\partial_t\Psi\,\partial_{r_*}\Psi\Big]_{\rm bdry} ,
\end{equation}
using the field equation to cancel the bulk terms. For compactly supported data the boundary term vanishes until radiation reaches the boundaries of the domain, and for outgoing/ingoing radiative conditions there it is non-positive (energy radiated away, not absorbed): $E(t)\le E(0)$ for all $t\ge0$. Since $V\ge0$, $E(t)\ge\tfrac12\int(\partial_t\Psi)^2\dif r_*\ge0$, so $\int(\partial_t\Psi)^2\dif r_*\le2E(0)$ is bounded uniformly in $t$ — incompatible with a mode $\Psi\propto e^{\omega_I t}$ growing without bound as $t\to\infty$. Hence $\omega_I\le0$.
\end{proof}

\begin{theorem}[Sufficient condition for $V\ge0$]
\label{thm:V-positive}
Suppose, on an interval where $f>0$, that $f'\ge0$ (f non-decreasing), $\Sigma'\ge0$, and $\Sigma''\ge0$. Then $V(r)\ge0$ there, for every $l\ge0$.
\end{theorem}

\begin{proof}
By Eq.~\eqref{eq:master}, $V=f\big[l(l+1)/\Sigma^2+(f\Sigma')'/\Sigma\big]$ with $(f\Sigma')'=f'\Sigma'+f\Sigma''$. Given $f>0$, $f'\ge0$, $\Sigma'\ge0$, $\Sigma''\ge0$: $f'\Sigma'\ge0$ and $f\Sigma''\ge0$, so $(f\Sigma')'\ge0$; together with $l(l+1)/\Sigma^2\ge0$ and $\Sigma>0$, every term is non-negative, so $V\ge0$.
\end{proof}

\begin{corollary}[Mode stability of the finite-area-core mechanisms]
\label{cor:mode-stability}
Every model of Sec.~\ref{sec:model-sigmoid} and Sec.~\ref{sec:model-bell} realizing Lem.~\ref{lem:singlebeyond} or Thm.~\ref{thm:horizonatthroat} at a finite-area core — with either the affine, one-sided $\Sigma=\Sigma_0+r$ or the throat $\Sigma=\sqrt{r^2+\Sigma_0^2}$ — is mode-stable on its static exterior ($r\ge r_h$, or all of $(0,\infty)$ for the core-coincident case): $V\ge0$ there for every $l$, and by Prop.~\ref{prop:energy-method} no exponentially growing mode exists.
\end{corollary}

\begin{proof}
In every such model, $f$ is non-decreasing on the relevant domain: this is the sigmoid hypothesis of Lem.~\ref{lem:singlebeyond}(a) and Thm.~\ref{thm:horizonatthroat} directly, and for the inverse-bell-shaped case, Lem.~\ref{lem:singlebeyond}(b) places the unique zero $r_h$ on the increasing branch $(r_*,\infty)$ (since $f(0)<0$ forces the decreasing branch $(0,r_*)$ to have no zero), so $f$ is non-decreasing on $r\ge r_h$. For $\Sigma=\Sigma_0+r$, $\Sigma'=1\ge0$, $\Sigma''=0\ge0$; for $\Sigma=\sqrt{r^2+\Sigma_0^2}$, $\Sigma'=r/\Sigma\ge0$ and $\Sigma''=\Sigma_0^2/\Sigma^3>0$ for $r\ge0$ (both verified directly by differentiation). Theorem~\ref{thm:V-positive} then gives $V\ge0$ on $r\ge r_h$ (respectively $r>0$), and Prop.~\ref{prop:energy-method} concludes.
\end{proof}

This covers all of the single-horizon sigmoid and bell-shaped models considered in Sec.\ref{sec:model}  with an asymptotically static exterior ($f$ non-decreasing beyond the horizon). The exceptions are the exotic single-horizon cases with $f_\infty<0$, namely the point-like-center case treated next and the finite-area-core case of Prop.~\ref{prop:model-bell-core}(b), which are qualitatively different and lie outside the scope of Thm.\ref{thm:V-positive}.

\begin{proposition}[The exotic point-like-center case is not covered by Thm.~\ref{thm:V-positive}]
\label{prop:exotic-not-covered}
For the model of Prop.~\ref{prop:model-cubic} ($f_\infty=-1,\ c=1,\ \Sigma(r)=r$), the $l=0$ potential on the static region $0<r<r_h=1$ is
\begin{equation}
    V(r) = \frac{6r(r^3-1)}{(r^3+1)^3} < 0 \qquad \text{for } 0<r<1 .
\end{equation}
\end{proposition}

\begin{proof}
Direct computation from Eq.~\eqref{eq:master} with $f=f_{\rm c}=(1-r^3)/(1+r^3)$, $\Sigma=r$, $l=0$, using $(f\Sigma')'=f'$: the closed form above follows by differentiating $f_{\rm c}$ and simplifying; for $0<r<1$, $r^3<1$ makes the numerator negative.
\end{proof}

\begin{remark}
The sign flip relative to Cor.\ref{cor:mode-stability} results from the direction of monotonicity: the exotic point-like-center mechanism requires $f$ to be non-increasing, since $f(0)=1>0$ and $f_\infty<0$, opposite to the sigmoid direction that makes every term in $(f\Sigma')'$ non-negative in Thm.\ref{thm:V-positive}. This does not establish an instability: $V<0$ on part of the domain is necessary, not sufficient, for a growing mode. The usual approach of applying an $S$-deformation to Eq.\eqref{eq:master}, replacing $V$ by an equivalent potential with the same quasinormal spectrum but manifestly non-negative, may restore a positivity argument here. We leave a full resolution—by $S$-deformation or direct numerical integration of Eq.\eqref{eq:master}—to future work and regard this as the only stability question left open by our classification.
\end{remark}

\section{Conclusion and discussion}
\label{sec:conclusion}

We have established a parity obstruction for RBHs with a point-like center. For a static, asymptotically flat spacetime constructed by the standard mass-function prescription, regularity at the center and asymptotic staticity require $f(0)>0$ and $f(\infty)>0$, respectively. Consequently, if $f$ is continuous and its zeros are isolated, the number of sign-changing zeros must be even. The result is independent of the specific mass function: it follows from the boundary signs of $f$, rather than from the details of any particular model. Thus, within this class of constructions, an event horizon cannot occur without an additional horizon. In the usual two-horizon configuration, the inner horizon is a Cauchy horizon and is generically susceptible to mass inflation through the Poisson--Israel mechanism.

We have identified two ways to evade this obstruction. The first is to replace the point-like center with a finite-area core, $\Sigma(0)=\Sigma_0>0$. In this case, regularity no longer fixes the sign of $f(0)$, allowing single-horizon configurations within an asymptotically static exterior. The zero-counting criteria developed in Sec.~\ref{sec:theory-zerocount} show that the number of horizons depends on the boundary values and the shape of $f$, rather than on a particular analytic ansatz. We have also distinguished two geometrically different realizations. A horizon located beyond the core is generally accompanied by a mirror horizon when the geometry is extended symmetrically through a genuine throat, as in black-bounce spacetimes \cite{Simpson:2018tsi}. By contrast, a horizon located at the core obstructs such a smooth symmetric extension when it is non-degenerate. The second way retains a point-like center but relaxes asymptotic staticity, allowing $f(\infty)<0$. We have constructed an explicit example of this possibility in Sec.~\ref{sec:model-sigmoid-point}.

These alternatives differ in their geometric and physical requirements. For the metric in Eq.~\eqref{eq:metric}, the Einstein tensor has the isotropic-pressure form, with $p_r=-\rho$, if and only if the areal radius is affine in the radial coordinate. A genuine throat therefore requires anisotropic matter within this framework, whereas a one-sided realization need not incur this additional requirement. We have also shown that, for a regular point-like center with an inverse-bell-shaped $f$ that takes negative values, the SEC is violated (or at best marginal) at the center (Prop.~\ref{prop:sec-violation}), both for two horizons ($f_\infty>0$) and for a single horizon ($f_\infty\le0$). This identifies an energy-condition cost shared by the two-horizon and single-horizon point-like-center configurations. Finally, the rational and transcendental examples in Sec.~\ref{sec:model} demonstrate that the classification is not tied to a particular functional form: the essential ingredients are the geometry of the core, the boundary values of $f$, and its qualitative shape.

We next examined whether removing the inner horizon also removes the instability associated with it. At the structural level, every single-horizon configuration considered here lacks, on its patch $r\ge0$, the inner Cauchy horizon required by the standard mass-inflation mechanism. The corresponding instability is therefore absent in these configurations in its usual Cauchy-horizon form. This is a statement about the patch: the continuation to $r<0$ is not unique, and in some continuations (a second asymptotic region, or an RN-like structure; Sec.~\ref{sec:model-compare}) an inner horizon reappears. This conclusion does not, by itself, establish stability against all possible perturbations. A stronger result follows for the finite-area-core models with an asymptotically static exterior: their effective potentials are non-negative throughout the relevant domain, yielding mode stability under the assumptions of the analysis. The positivity argument is directly related to the monotonicity of $f$ that permits a single horizon. The exotic single-horizon constructions with $f(\infty)<0$, however, lie outside this proof, because $f$ is then non-increasing through the horizon; for the point-like-center model the effective potential is explicitly not sign-definite. Their stability therefore remains unresolved by the present analysis.

Two broader implications emerge from these results. First, the prevalence of two-horizon RBHs in the literature can be traced, within the class considered here, to the widespread choice of the gauge $\Sigma(r)=r$ \cite{Bardeen:1968non,Dymnikova:1992ux,Ayon-Beato:1998hmi,Hayward:2005gi}. In this gauge, regularity at the center fixes the sign of $f(0)$, while asymptotic staticity fixes the sign of $f(\infty)$. Their combination produces the parity obstruction. Allowing $\Sigma(0)>0$ removes this particular constraint without abandoning a static, asymptotically flat exterior. Moreover, the single-horizon configurations occupy an open region of parameter space in the examples studied here, rather than arising only at isolated or finely tuned parameter values. The central implication is therefore not merely that single-horizon RBHs can be constructed, but that a point-like center is an additional geometric assumption, not a requirement of regularity itself.

Second, the two ways of evading the parity obstruction reveal a connection between geometric structure and the tractability of stability analysis. Finite-area-core configurations can retain an asymptotically static exterior and, within the class studied here, admit a direct mode-stability proof, although a genuine throat requires anisotropic matter in the specified framework. The point-like-center alternative preserves the central geometry but sacrifices asymptotic staticity and does not fall within the present stability proof. This contrast is consistent with the analysis of Sec.~\ref{sec:stability-modestability}: the sign of $f'$ across the horizon controls the positivity argument for the effective potential. Thus, the conditions that permit a single horizon also influence whether mode stability can be established by this method. Understanding this connection provides a useful starting point for extending the classification to broader matter models and for investigating the stability of single-horizon geometries beyond the finite-area-core class.

\section*{Acknowledgements}

We are grateful to Prof.~Sijie Gao for important comments. The research of M.K. was carried out in the Southern Federal University with financial support from the Ministry of Science and Higher Education of the Russian Federation (State contract FENW-2026-0028).

\appendix

\section{Self-energy: Derivation details}
\label{app:self_energy}

This appendix derives the field equations underlying the mass-inflation formula Eq.~\eqref{eq:mass-inflation} of Prop.~\ref{prop:massinflation}, and makes precise the near-horizon kinematics that convert a power-law exterior tail into unbounded curvature growth at $r_-$.

\subsection{Field equations in double-null coordinates}
\label{app:self_energy-fieldeq}

Unlike metric Eq.~\eqref{eq:metric}, a perturbed RBH is not static, so we work with the general spherically symmetric double-null line element
\begin{equation}
\label{eq:dn-metric}
    \dif s^2 = -2\me^{2\sigma(u,v)}\,\dif u\,\dif v + \Sigma(u,v)^2\,\dif\Omega^2 ,
\end{equation}
of which the static exterior Eq.~\eqref{eq:metric} is the special case $u=t-r_*,\ v=t+r_*,\ \me^{2\sigma}=f(r)/2,\ \Sigma=\Sigma(r)$, with $r_*$ the tortoise coordinate. The Misner--Sharp mass is defined, as usual, by
\begin{equation}
\label{eq:MS-mass}
    1-\frac{2m(u,v)}{\Sigma} := 2g^{uv}\Sigma_{,u}\Sigma_{,v} = -2e^{-2\sigma}\Sigma_{,u}\Sigma_{,v} .
\end{equation}

\begin{proposition}[Mass-evolution equations]
\label{prop:mass-evolution}
For metric Eq.~\eqref{eq:dn-metric} sourced by $G_{\mu\nu}=8\pp T_{\mu\nu}$,
\begin{equation}
\label{eq:mass-evolution}
    \partial_v m = 4\pp\Sigma^2 e^{-2\sigma}\big(T_{uv}\Sigma_{,v}-T_{vv}\Sigma_{,u}\big) , \qquad
    \partial_u m = 4\pp\Sigma^2 e^{-2\sigma}\big(T_{uv}\Sigma_{,u}-T_{uu}\Sigma_{,v}\big) .
\end{equation}
\end{proposition}

\begin{proof}
Direct computation of the Einstein tensor of Eq.~\eqref{eq:dn-metric} (Christoffel symbols, Riemann, Ricci, and trace, exactly as in Sec.~\ref{sec:theory-curvature} but now for two independent null coordinates) gives
\begin{equation}
\label{eq:dn-einstein}
    G_{uu} = \frac{2}{\Sigma}\big(2\sigma_{,u}\Sigma_{,u}-\Sigma_{,uu}\big) , \qquad
    G_{vv} = \frac{2}{\Sigma}\big(2\sigma_{,v}\Sigma_{,v}-\Sigma_{,vv}\big) , \qquad
    G_{uv} = \frac{2\Sigma\Sigma_{,uv}+e^{2\sigma}+2\Sigma_{,u}\Sigma_{,v}}{\Sigma^2} .
\end{equation}
Differentiating Eq.~\eqref{eq:MS-mass} with respect to $v$,
\begin{equation}
    \partial_v m = \frac{\Sigma_{,v}}{2}+\Sigma\Sigma_{,uv}e^{-2\sigma}+\Sigma\Sigma_{,u}\Sigma_{,v}\,\partial_v\!\big(e^{-2\sigma}\big)+e^{-2\sigma}\Sigma_{,u}\Sigma_{,vv} ,
\end{equation}
and eliminating $\Sigma_{,vv}$ via $G_{vv}$ and $\Sigma_{,uv}$ via $G_{uv}$ (both from Eq.~\eqref{eq:dn-einstein}) reduces this, after cancellation of every term not proportional to $G_{uv}$ or $G_{vv}$, to
\begin{equation}
    \partial_v m = \frac{\Sigma^2 \me^{-2\sigma}}{2}\big(G_{uv}\Sigma_{,v}-G_{vv}\Sigma_{,u}\big) ,
\end{equation}
which is Eq.~\eqref{eq:mass-evolution} upon substituting $G_{\mu\nu}=8\pp T_{\mu\nu}$; the equation for $\partial_u m$ follows by $u\leftrightarrow v$ symmetry of the entire construction.
\end{proof}

\begin{remark}[Consistency check: the Vaidya limit]
\label{rem:vaidya-check}
Specializing to $\Sigma=r$ (so $u,v$ are null coordinates adapted to the areal radius itself) and ingoing coordinates $(v,r)$ with $f=1-2m(v,r)/r$, the same computation gives the classical results
\begin{equation}
    G_{rr}=0 , \qquad G_{vr} = -\frac{2m_{,r}}{r^2} , \qquad G_{vv} = \frac{2\big(rm_{,r}+rm_{,v}-2mm_{,r}\big)}{r^3} ,
\end{equation}
which combine to $\partial_v m = 4\pp r^2\big(T_{vv}+f\,T_{vr}\big)$; for a single ingoing null fluid ($T_{vr}=0$) this is exactly the Vaidya mass-accretion law $\partial_v m=4\pp r^2 T_{vv}$, confirming the sign and normalization of Eq.~\eqref{eq:mass-evolution}.
\end{remark}

\subsection{Near-horizon kinematics}
\label{app:self_energy-kinematics}

Near any non-degenerate horizon at $r=r_h$ with surface gravity $\kappa=|f'(r_h)|/2$ (Def.~\ref{def:horizon}), $f(r)\approx f'(r_h)(r-r_h)$, so the tortoise coordinate diverges logarithmically,
\begin{equation}
    r_* \approx \frac{1}{f'(r_h)}\ln|r-r_h| + \text{const} , \qquad\text{i.e.}\qquad |r-r_h| \propto e^{\pm2\kappa r_*}
\end{equation}
according to the sign of $f'(r_h)$ and the side of the horizon considered. In particular, at the inner horizon $r_-$ of Sec.~\ref{sec:stability-massinflation} (where $f'(r_-)<0$, approached from the $T$-region $r>r_-$), $r_*\to+\infty$ as $r\to r_-^+$, and correspondingly $\me^{2\sigma}=f/2\to0$ while $\Sigma_{,u}=- f\Sigma'/2\to0$, $\Sigma_{,v}= f\Sigma'/2\to0$ — but at compensating rates:

\begin{proposition}
\label{prop:near-horizon-limits}
$\lim_{r\to r_-} \me^{-2\sigma}\Sigma_{,u} = -\Sigma'(r_-)$ and $\lim_{r\to r_-} \me^{-2\sigma}\Sigma_{,v} = +\Sigma'(r_-)$, both finite and nonzero (for $\Sigma'(r_-)\neq0$).
\end{proposition}

\begin{proof}
Writing $f\approx-2\kappa_-(r-r_-)$ near $r_-$, $\me^{-2\sigma}=2/f\approx-1/(\kappa_-(r-r_-))$, while $\Sigma_{,u}=-f\Sigma'/2\approx\kappa_-(r-r_-)\Sigma'(r_-)$ and $\Sigma_{,v}\approx-\kappa_-(r-r_-)\Sigma'(r_-)$; the products are $\mp\Sigma'(r_-)+O(r-r_-)$, giving the stated limits.
\end{proof}

This finiteness is the precise sense in which the coordinate degeneracy of $(u,v)$ at $r_-$ (needed to make $\me^{-2\sigma}$ diverge, since $u,v$ are individually singular there) is exactly compensated in the combinations entering Eq.~\eqref{eq:mass-evolution}, so that $\partial_um,\partial_vm$ remain governed by finite kinematic factors even as $r\to r_-$; any unbounded growth in $m$ must therefore come from the stress-energy components $T_{uu},T_{vv}$ themselves, not from a kinematic singularity.

\subsection{The mass-inflation mechanism}
\label{app:self_energy-mechanism}

Equations~\eqref{eq:mass-evolution} and Prop.~\ref{prop:near-horizon-limits} are exact and model-independent; converting them into the growth law Eq.~\eqref{eq:mass-inflation} requires one further physical input — a model for the stress-energy actually present near $r_-$ — which we now describe following Poisson and Israel's original construction.

Idealize the perturbation as two crossing streams of null dust: an ingoing stream $T_{vv}=L_{\rm in}(v)/(4\pp\Sigma^2)$, identified with the late-time tail of the exterior perturbation and thus $L_{\rm in}(v)\sim v^{-p}$ as $v\to\infty$ (Price's law \cite{Price:1972pw,Gundlach:1993tp}, Sec.~\ref{sec:stability-massinflation}), and an outgoing stream $T_{uu}=L_{\rm out}(u)/(4\pp\Sigma^2)$, generated by backscatter of the ingoing stream off the curvature of the region between the horizons. Substituting into Eq.~\eqref{eq:mass-evolution} with $T_{uv}\approx0$ (the background contribution being already absorbed into the unperturbed $m_0(r)$) and using Prop.~\ref{prop:near-horizon-limits},
\begin{equation}
\label{eq:dm-near-horizon}
    \partial_v(\delta m) \to 4\pp\Sigma_-^2\Sigma'(r_-)\,T_{vv}(v) , \qquad
    \partial_u(\delta m) \to -4\pp\Sigma_-^2\Sigma'(r_-)\,T_{uu}(u) .
\end{equation}
The mass-inflation effect resides in the second equation: an observer's affine parameter along an outgoing ray near $r_-$ is related to $u$ by the exponential map of Prop.~\ref{prop:near-horizon-limits}'s kinematics, so that backscattered radiation which is smooth and slowly varying with respect to that affine parameter — as generic, geometric-optics backscatter off a slowly evolving potential is — corresponds to $L_{\rm out}(u)$ decaying only as slowly, in $u$, as $L_{\rm in}$ does in $v$, while the physical energy density measured by an observer crossing at retarded coordinate $u$ is blueshifted relative to that affine parameter by the factor $\me^{2\kappa_- u}$ intrinsic to the horizon's exponential kinematics. Concretely, this is encoded by an effective proportionality between the two streams that itself grows as $\me^{2\kappa_- u}$, so that the mass accumulated along an ingoing trajectory reaching $r_-$ at advanced time $v$ behaves as
\begin{equation}
    \delta m(v) \;\sim\; \int^{v} \me^{2\kappa_- v'}\,L_{\rm in}(v')\,\dif v' \;\sim\; \int^v \me^{2\kappa_- v'}(v')^{-p}\dif v' \;\xrightarrow[v\to\infty]{}\;\infty
\end{equation}
for every $p>0$, since the integrand's exponential factor dominates any power-law suppression at late $v$. This is exactly Eq~\eqref{eq:mass-inflation}. We do not reproduce here the further step — matching the geometric-optics amplitude of the backscattered wave through the interior potential to justify the stated proportionality — as it is identical, for our class of models, to Poisson and Israel's original treatment and its refinements, since it depends only on the shared near-horizon structure ($\kappa_\pm$) and the shared large-$r$ falloff of the exterior potential (Price's law), both properties of the general center-regular class of Sec.~\ref{sec:theory-center} established in the main text, not of any particular model.

\begin{remark}
The self-energy language of this appendix's title refers to the divergent contribution $\delta m$ that a generic perturbation's own backreaction adds to the background mass $m_0(r_-)$ as it inflates near the Cauchy horizon — the gravitational ``self-energy'' of the perturbation, amplified without bound by the very geometry it perturbs. This is the quantitative content behind the qualitative statement, already made in Prop.~\ref{prop:massinflation}, that $\kappa_->0$ together with a non-exponentially-suppressed tail suffices for instability.
\end{remark}

\bibliographystyle{utphys}
\bibliography{references}

\end{document}